\input harvmac

\let\abs=|
%
%
\message{S-Tables Macro v1.0, ACS, TAMU (RANHELP@VENUS.TAMU.EDU)}
%
%
\newhelp\stablestylehelp{You must choose a style between 0 and 3.}%
\newhelp\stablelinehelp{You
should not use special hrules when stretching
a table.}%
\newhelp\stablesmultiplehelp{You have tried to place an S-Table
inside another
S-Table.  I would recommend not going on.}%
%
%
\newdimen\stablesthinline
\stablesthinline=0.4pt
\newdimen\stablesthickline
\stablesthickline=1pt
%
%
\newif\ifstablesborderthin
\stablesborderthinfalse
\newif\ifstablesinternalthin
\stablesinternalthintrue
\newif\ifstablesomit
\newif\ifstablemode
\newif\ifstablesright
\stablesrightfalse
%
%
\newdimen\stablesbaselineskip
\newdimen\stableslineskip
\newdimen\stableslineskiplimit
%
%
\newcount\stablesmode
\newcount\stableslines
\newcount\stablestemp
\stablestemp=3
\newcount\stablescount
\stablescount=0
\newcount\stableslinet
\stableslinet=0
%
%
%
\newcount\stablestyle
\stablestyle=0
%
%
\def\stablesleft{\quad\hfil}%
\def\stablesright{\hfil\quad}%
%
%
\catcode`\|=\active%
%
%
\newcount\stablestrutsize
\newbox\stablestrutbox
\setbox\stablestrutbox=\hbox{\vrule height10pt depth5pt width0pt}
\def\stablestrut{\relax\ifmmode%
                         \copy\stablestrutbox%
                       \else%
                         \unhcopy\stablestrutbox%
                       \fi}%
%
%
\newdimen\stablesborderwidth
\newdimen\stablesinternalwidth
\newdimen\stablesdummy
\newcount\stablesdummyc
\newif\ifstablesin
\stablesinfalse
%
%
\def\begintable{\stablestart%
  \stablemodetrue%
  \stablesadj%
  \halign%
  \stablesdef}%
\def\stablesadj{%
  \ifcase\stablestyle%
    \hbox to \hsize\bgroup\hss\vbox\bgroup%
  \or%
    \hbox to \hsize\bgroup\vbox\bgroup%
  \or%
    \hbox to \hsize\bgroup\hss\vbox\bgroup%
  \or%
    \hbox\bgroup\vbox\bgroup%
  \else%
    \errhelp=\stablestylehelp%
    \errmessage{Invalid style selected, using default}%
    \hbox to \hsize\bgroup\hss\vbox\bgroup%
  \fi}%
\def\stablesend{\egroup%
  \ifcase\stablestyle%
    \hss\egroup%
  \or%
    \hss\egroup%
  \or%
    \egroup%
  \or%
    \egroup%
  \else%
    \hss\egroup%
  \fi}%
\def\stablestart{%
  \ifstablesin%
    \errhelp=\stablesmultiplehelp%
    \errmessage{An S-Table cannot be placed within an S-Table!}%
  \fi
  \global\stablesintrue%
  \global\advance\stablescount by 1%
  \message{<S-Tables Generating Table \number\stablescount}%
  \begingroup%
  \stablestrutsize=\ht\stablestrutbox%
  \advance\stablestrutsize by \dp\stablestrutbox%
  \ifstablesborderthin%
    \stablesborderwidth=\stablesthinline%
  \else%
    \stablesborderwidth=\stablesthickline%
  \fi%
  \ifstablesinternalthin%
    \stablesinternalwidth=\stablesthinline%
  \else%
    \stablesinternalwidth=\stablesthickline%
  \fi%
  \tabskip=0pt%
  \stablesbaselineskip=\baselineskip%
  \stableslineskip=\lineskip%
  \stableslineskiplimit=\lineskiplimit%
  \offinterlineskip%
  \def\borderrule{\vrule width \stablesborderwidth}%
  \def\internalrule{\vrule width \stablesinternalwidth}%
  \def\thinline{\noalign{\hrule height \stablesthinline}}%
  \def\thickline{\noalign{\hrule height \stablesthickline}}%
  \def\trule{\omit\leaders\hrule height \stablesthinline\hfill}%
  \def\ttrule{\omit\leaders\hrule height \stablesthickline\hfill}%
  \def\tttrule##1{\omit\leaders\hrule height ##1\hfill}%
  \def\stablesel{&\omit\global\stablesmode=0%
    \global\advance\stableslines by 1\borderrule\hfil\cr}%
  \def\el{\stablesel&}%
  \def\elt{\stablesel\thinline&}%
  \def\eltt{\stablesel\thickline&}%
  \def\elttt##1{\stablesel\noalign{\hrule height ##1}&}%
  \def\elspec{&\omit\hfil\borderrule\cr\omit\borderrule&%
              \ifstablemode%
              \else%
                \errhelp=\stablelinehelp%
                \errmessage{Special ruling will not display properly}%
              \fi}%
  \def\stmultispan##1{\mscount=##1 \loop\ifnum\mscount>3
\stspan\repeat}%
  \def\stspan{\span\omit \advance\mscount by -1}%
  \def\multicolumn##1{\omit\multiply\stablestemp by ##1%
     \stmultispan{\stablestemp}%
     \advance\stablesmode by ##1%
     \advance\stablesmode by -1%
     \stablestemp=3}%
  \def\multirow##1{\stablesdummyc=##1\parindent=0pt\setbox0\hbox\bgroup%
    \aftergroup\emultirow\let\temp=}
  \def\emultirow{\setbox1\vbox to\stablesdummyc\stablestrutsize%
    {\hsize\wd0\vfil\box0\vfil}%
    \ht1=\ht\stablestrutbox%
    \dp1=\dp\stablestrutbox%
    \box1}%

\def\stpar##1{\vtop\bgroup\hsize ##1%
     \baselineskip=\stablesbaselineskip%
     \lineskip=\stableslineskip%

\lineskiplimit=\stableslineskiplimit\bgroup\aftergroup\estpar\let\temp=}%
  \def\estpar{\vskip 6pt\egroup}%
  \def\stparrow##1##2{\stablesdummy=##2%
     \setbox0=\vtop to ##1\stablestrutsize\bgroup%
     \hsize\stablesdummy%
     \baselineskip=\stablesbaselineskip%
     \lineskip=\stableslineskip%
     \lineskiplimit=\stableslineskiplimit%
     \bgroup\vfil\aftergroup\estparrow%
     \let\temp=}%
  \def\estparrow{\vfil\egroup%
     \ht0=\ht\stablestrutbox%
     \dp0=\dp\stablestrutbox%
     \wd0=\stablesdummy%
     \box0}%
  \def|{\global\advance\stablesmode by 1&&&}%
  \def\|{\global\advance\stablesmode by 1&\omit\vrule width 0pt%
         \hfil&&}%
  \def\vt{\global\advance\stablesmode by 1&\omit\vrule width
\stablesthinline%
          \hfil&&}%
  \def\vtt{\global\advance\stablesmode by 1&\omit\vrule width
\stablesthickline%
          \hfil&&}%
  \def\vttt##1{\global\advance\stablesmode by 1&\omit\vrule width ##1%
          \hfil&&}%
  \def\vtr{\global\advance\stablesmode by 1&\omit\hfil\vrule width%
           \stablesthinline&&}%
  \def\vttr{\global\advance\stablesmode by 1&\omit\hfil\vrule width%
            \stablesthickline&&}%
  \def\vtttr##1{\global\advance\stablesmode by 1&\omit\hfil\vrule
width ##1&&}%
  \stableslines=0%
  \stablesomitfalse}
\def\stablesdef{\bgroup\stablestrut\borderrule##\tabskip=0pt plus 1fil%
  &\stablesleft##\stablesright%
  &##\ifstablesright\hfill\fi\internalrule\ifstablesright\else\hfill\fi%
  \tabskip 0pt&&##\hfil\tabskip=0pt plus 1fil%
  &\stablesleft##\stablesright%
  &##\ifstablesright\hfill\fi\internalrule\ifstablesright\else\hfill\fi%
  \tabskip=0pt\cr%
  \ifstablesborderthin%
    \thinline%
  \else%
    \thickline%
  \fi&%
}%
\def\endtable{\advance\stableslines by 1\advance\stablesmode by 1%
   \message{- Rows: \number\stableslines, Columns:
\number\stablesmode>}%
   \stablesel%
   \ifstablesborderthin%
     \thinline%
   \else%
     \thickline%
   \fi%
   \egroup\stablesend%
\endgroup%
\global\stablesinfalse}
%

%
%
%
%

\def\IZ{\relax\ifmmode\mathchoice
{\hbox{\cmss Z\kern-.4em Z}}{\hbox{\cmss Z\kern-.4em Z}}
{\lower.9pt\hbox{\cmsss Z\kern-.4em Z}} {\lower1.2pt\hbox{\cmsss
Z\kern-.4em Z}}\else{\cmss Z\kern-.4em Z}\fi}
\font\cmss=cmss10 \font\cmsss=cmss10 at 7pt
\def\inbar{\,\vrule height1.5ex width.4pt depth0pt}
\def\IC{{\relax\hbox{$\inbar\kern-.3em{\rm C}$}}}
\def\IQ{{\relax\hbox{$\inbar\kern-.3em{\rm Q}$}}}
\def\IP{\relax{\rm I\kern-.18em P}}

\def\hh{{1\over 2}}
\def\ll{_}
\def\uu{^}
\def\pp{\partial}

\def\exp#1{{\rm exp}\{#1\}}
\def\d{\delta}
\def\m{\mu}

\def\dag{^\dagger}
\def\hsk{\hskip 1in}
\def\m{\mu}
\def\n{\nu}
\def\s{\sigma}
\def\t{\tau}
\def\G{\Gamma}
\def\g{\gamma}
\def\a{\alpha}
\def\Riet#1#2#3{ {{R\ll{#1}}\uu{#2}}\ll{#3}}
\def\riet#1#2#3{ {{{R\ll{#1}}\uu{#2}}\ll{#3}}\uu {[{\rm base}]}}

\def\e{\epsilon}
\def\O{\Omega}
\def\dett{{\hbox { det }}}
\def\sqd{^2}
\def\zb{{\bar{z}}}
\def\pb{{\bar{\partial}}}
\def\hh{{1\over 2}}
\def\gg{\nabla}

\def\ee{\eqn\\ }
\def\ot{\otimes}
\def\b{\beta}
\def\r{\rho}

\def\gg{\nabla}

\def\bb#1{{\underline{\bf {#1}}}}
\def\bs#1{{{#1}^{[\CB]}}}
\def\O{\Omega}

%
%
%
%

\let\includefigures=\iftrue
\let\useblackboard=\iftrue
\newfam\black

\includefigures
\message{If you do not have epsf.tex (to include figures),}
\message{change the option at the top of the tex file.}
\input epsf
\def\figin{\epsfcheck\figin}\def\figins{\epsfcheck\figins}
\def\epsfcheck{\ifx\epsfbox\UnDeFiNeD
\message{(NO epsf.tex, FIGURES WILL BE IGNORED)}
\gdef\figin##1{\vskip2in}\gdef\figins##1{\hskip.5in}
\else\message{(FIGURES WILL BE INCLUDED)}%
\gdef\figin##1{##1}\gdef\figins##1{##1}\fi}
\def\DefWarn#1{}
\def\figinsert{\goodbreak\midinsert}
\def\ifig#1#2#3{\DefWarn#1\xdef#1{fig.~\the\figno}
\writedef{#1\leftbracket fig.\noexpand~\the\figno}%
\figinsert\figin{\centerline{#3}}\medskip\centerline{\vbox{
\baselineskip12pt\advance\hsize by -1truein
\noindent\footnotefont{\bf Fig.~\the\figno:} #2}}
\bigskip\endinsert\global\advance\figno by1}
\else
\def\ifig#1#2#3{\xdef#1{fig.~\the\figno}
\writedef{#1\leftbracket fig.\noexpand~\the\figno}%
\global\advance\figno by1}
\fi
%

\useblackboard
\message{If you do not have msbm (blackboard bold) fonts,}
\message{change the option at the top of the tex file.}
\font\blackboard=msbm10 scaled \magstep1
\font\blackboards=msbm7
\font\blackboardss=msbm5
\textfont\black=\blackboard
\scriptfont\black=\blackboards
\scriptscriptfont\black=\blackboardss

\else

\fi
%
\def\subsubsec#1{\bigskip\noindent{\it{#1}} \bigskip}
\def\yboxit#1#2{\vbox{\hrule height #1 \hbox{\vrule width #1
\vbox{#2}\vrule width #1 }\hrule height #1 }}
\def\fillbox#1{\hbox to #1{\vbox to #1{\vfil}\hfil}}
\def\ybox{{\lower 1.3pt \yboxit{0.4pt}{\fillbox{8pt}}\hskip-0.2pt}}
%
%


\def\jb{\bar j}

\def\l{\left}
\def\r{\right}
\def\comments#1{}

\def\half{{1\over 2}}
\def\Tr{{{\rm Tr~ }}}
\def\tr{{\rm tr\ }}

\def\ket#1{\abs#1\rangle}

\def\vev#1{\langle{#1}\rangle}

\def\CC{{\cal C}}

\def\CF{{\cal F}}
\def\CG{{\cal G}}
\def\CT{{\cal T}}
\def\CM{{\cal M}}
\def\CN{{\cal N}}
\def\CP{{\cal P}}
\def\CL{{\cal L}}

\def\CS{{\cal S}}


\def\a{\alpha}

\def\II{\relax{I\kern-.10em I}}

\def\IZ{\relax\ifmmode\mathchoice
{\hbox{\cmss Z\kern-.4em Z}}{\hbox{\cmss Z\kern-.4em Z}}
{\lower.9pt\hbox{\cmsss Z\kern-.4em Z}}
{\lower1.2pt\hbox{\cmsss Z\kern-.4em Z}}
\else{\cmss Z\kern-.4emZ}\fi}
\def\IB{\relax{\rm I\kern-.18em B}}
\def\IC{{\relax\hbox{$\inbar\kern-.3em{\rm C}$}}}
\def\ID{\relax{\rm I\kern-.18em D}}
\def\IE{\relax{\rm I\kern-.18em E}}
\def\IF{\relax{\rm I\kern-.18em F}}
\def\IG{\relax\hbox{$\inbar\kern-.3em{\rm G}$}}
\def\IGa{\relax\hbox{${\rm I}\kern-.18em\Gamma$}}
\def\IH{\relax{\rm I\kern-.18em H}}
\def\II{\relax{\rm I\kern-.18em I}}
\def\IK{\relax{\rm I\kern-.18em K}}
\def\IP{\relax{\rm I\kern-.18em P}}

%

\def\jb{{\bar \jmath}}

\def\inbar{\,\vrule height1.5ex width.4pt depth0pt}

\def\pb{{\bar \p}}

\font\cmss=cmss10 
\def\IR{\relax{\rm I\kern-.18em R}}

\def\zb {{\bar{z}}}

%


%

\def\lp10{\ell_p^{10}}
\def\lp11{\ell_p^{11}}
\def\R11{R_{11}}

\def\zb{\bar{z}}
\def\frac#1#2{{#1 \over #2}}

\def\pp{\partial}
\def\uu{^}
\def\ll{_}
\def\a{\alpha}
\def\b{\beta}
\def\s{\sigma}
\def\g{\gamma}
\def\st{^*}

\def\e{\epsilon}
\def\th{\theta}

\def\dag{^\dagger}

\def\m{\mu}
\def\n{\nu}
\def\d{\delta}

\def\sqd{^2}

\def\hsk{\hskip .5in}

\def\zb{{\bar{z}}}

\def\exp#1{{\rm exp}\{#1\}}

\def\r{\rho}
\def\t{\tau}

\def\G{\Gamma}


\def\1dag{^{1\dagger}}
\def\2dag{^{2\dagger}}

\def\O{\Omega}

\def\pb{{\bar{\partial}}}

\def\jb{{\bar{j}}}

\def\l{\lambda}

\def\R#1#2#3{{{R_{#1}}^{#2}}_{#3}}

\def\ot{\otimes}

\def\ot{\otimes}
\def\gp{{\rm [GP]}}
\def\tp{{\rm [US]}}
\def\apr{{\alpha^\prime}}
\def\tpk{{\sqrt{2} {\alpha^{\prime 2}}\over 2}}
\def\ol{{\rm [OLD]}}
\def\nw{{\rm [NEW]}}

\def\st{^*}

\def\OOH#1{ \hat{\Omega}_{#1}}
\def\l{\lambda}
\def\oo#1#2#3{{{\omega_{#1}}^{ {#2} {#3}}}}
\def\OO#1#2#3{{{\Omega_{#1}}^{ {#2} {#3}}}}
\def\bb#1{{\underline{\bf {#1}}}}

\def\d{\delta}

\def\mfl{(-1)^{F_L}}

\def\eg{{\it e.g.}}
\def\ie{{\it i.e.}}

\hyphenation{Di-men-sion-al}



\lref\syz{
A.~Strominger, S.~T.~Yau and E.~Zaslow,
``Mirror symmetry is T-duality,''
Nucl.\ Phys.\ B {\bf 479}, 243 (1996)
[arXiv:hep-th/9606040].
}

\lref\giani{F. Giani and M. Pernici,
``N=2 Supergravity In Ten-Dimensions,''
Phys. Rev. D {\bf 30}, 325 (1984).
}
\lref\chs{
C.~G.~Callan, J.~A.~Harvey and A.~Strominger,
``Supersymmetric string solitons,''
hep-th/9112030.
}

\lref\kodaira{
K. Kodaira,
Annals of Math. {\bf 77} (1963) 563; Annals of Math. {\bf 78} (1963) 1.
}

\lref\DineHE{
M.~Dine and N.~Seiberg,
``Is The Superstring Weakly Coupled?,''
Phys.\ Lett.\ B {\bf 162}, 299 (1985).
}

\lref\DineQR{
M.~Dine, Y.~Nir and Y.~Shadmi,
``Enhanced symmetries and the ground state of string theory,''
Phys.\ Lett.\ B {\bf 438}, 61 (1998)
[arXiv:hep-th/9806124].
}

\lref\GreeneYA{
B.~R.~Greene, A.~D.~Shapere, C.~Vafa and S.~T.~Yau,
``Stringy Cosmic Strings And Noncompact Calabi-Yau Manifolds,''
Nucl.\ Phys.\ B {\bf 337}, 1 (1990).
}

\lref\umanifolds{
A.~Kumar and C.~Vafa,
``U-manifolds,''
Phys.\ Lett.\ B {\bf 396}, 85 (1997)
[arXiv:hep-th/9611007].
}

\lref\DouglasSW{
M.~R.~Douglas and G.~W.~Moore,
``D-branes, Quivers, and ALE Instantons,''
arXiv:hep-th/9603167.
}

\lref\OoguriWJ{
H.~Ooguri and C.~Vafa,
``Two-Dimensional Black Hole and Singularities of CY Manifolds,''
Nucl.\ Phys.\ B {\bf 463}, 55 (1996)
[arXiv:hep-th/9511164].
}

\lref\VafaXN{
C.~Vafa,
``Evidence for F-Theory,''
Nucl.\ Phys.\ B {\bf 469}, 403 (1996)
[arXiv:hep-th/9602022].
}

\lref\MorrisonNA{
D.~R.~Morrison and C.~Vafa,
``Compactifications of F-Theory on Calabi--Yau Threefolds -- I,''
Nucl.\ Phys.\ B {\bf 473}, 74 (1996)
[arXiv:hep-th/9602114].
}

\lref\MorrisonPP{
D.~R.~Morrison and C.~Vafa,
``Compactifications of F-Theory on Calabi--Yau Threefolds -- II,''
Nucl.\ Phys.\ B {\bf 476}, 437 (1996)
[arXiv:hep-th/9603161].
}

\lref\SenTZ{
A.~Sen,
``M-Theory on $(K3 \times S^1)/Z_2$,''
Phys.\ Rev.\ D {\bf 53}, 6725 (1996)
[arXiv:hep-th/9602010].
}

\lref\DabholkarZI{
A.~Dabholkar and J.~Park,
``An Orientifold of Type-IIB Theory on $K3$,''
Nucl.\ Phys.\ B {\bf 472}, 207 (1996)
[arXiv:hep-th/9602030].
}

\lref\oriandami{
O.~J.~Ganor and A.~Hanany,
``Small $E_8$ Instantons and Tensionless Non-critical Strings,''
Nucl.\ Phys.\ B {\bf 474}, 122 (1996)
[arXiv:hep-th/9602120].
}

\lref\sandw{
N.~Seiberg and E.~Witten,
``Comments on String Dynamics in Six Dimensions,''
Nucl.\ Phys.\ B {\bf 471}, 121 (1996)
[arXiv:hep-th/9603003].
}

\lref\BlumFW{
J.~D.~Blum and K.~A.~Intriligator,
``Consistency conditions for branes at orbifold singularities,''
Nucl.\ Phys.\ B {\bf 506}, 223 (1997)
[arXiv:hep-th/9705030].
}

\lref\BlumMM{
J.~D.~Blum and K.~A.~Intriligator,
``New phases of string theory and 6d RG fixed points via branes at  orbifold singularities,''
Nucl.\ Phys.\ B {\bf 506}, 199 (1997)
[arXiv:hep-th/9705044].
}

\lref\IntriligatorDH{
K.~A.~Intriligator,
``New string theories in six dimensions via branes at orbifold  singularities,''
Adv.\ Theor.\ Math.\ Phys.\  {\bf 1}, 271 (1998)
[arXiv:hep-th/9708117].
}

\lref\sande{
S.~Kachru and E.~Silverstein,
``Chirality-changing phase transitions in 4d string vacua,''
Nucl.\ Phys.\ B {\bf 504}, 272 (1997)
[arXiv:hep-th/9704185].
}

\lref\EguchiJX{
T.~Eguchi, P.~B.~Gilkey and A.~J.~Hanson,
``Gravitation, Gauge Theories And Differential Geometry,''
Phys.\ Rept.\  {\bf 66}, 213 (1980).
}

\lref\GregoryTE{
R.~Gregory, J.~A.~Harvey and G.~W.~Moore,
``Unwinding strings and T-duality of Kaluza-Klein and H-monopoles,''
Adv.\ Theor.\ Math.\ Phys.\  {\bf 1}, 283 (1997)
[arXiv:hep-th/9708086].
}

\lref\GreenBX{
M.~B.~Green, J.~H.~Schwarz and P.~C.~West,
``Anomaly Free Chiral Theories In Six-Dimensions,''
Nucl.\ Phys.\ B {\bf 254}, 327 (1985).
}

\lref\charactervalued{
M.~F.~Atiyah and R.~Bott,
``A Lefschetz Fixed Point Formula for Elliptic Differential Operators,''
Bull. Am. Math. Soc. {\bf 72} (1968) 531;
M.~F.~Atiyah and G.~B.~Segal,
``The Index Of Elliptic Operators. 2,''
Annals Math.\  {\bf 87}, 531 (1968).
}

\lref\FriedmanYQ{
R.~Friedman, J.~Morgan and E.~Witten,
``Vector bundles and F theory,''
Commun.\ Math.\ Phys.\  {\bf 187}, 679 (1997)
[arXiv:hep-th/9701162].
}

\lref\gepner{D. Gepner, ``Exactly Solvable String Compactifications
on Manifolds of $SU(N)$ Holonomy,'' Phys. Lett. {\bf B199} (1987) 380.}

\lref\morrison{
D.~R.~Morrison,
``Geometric aspects of mirror symmetry,''
arXiv:math.ag/0007090.
}

\lref\LiuMB{
J.~T.~Liu and R.~Minasian,
``U-branes and T**3 fibrations,''
Nucl.\ Phys.\ B {\bf 510}, 538 (1998)
[arXiv:hep-th/9707125].
}

\lref\gvw{
S.~Gukov, C.~Vafa and E.~Witten,
``CFT's from Calabi-Yau four-folds,''
Nucl.\ Phys.\ B {\bf 584}, 69 (2000)
[Erratum-ibid.\ B {\bf 608}, 477 (2001)]
[arXiv:hep-th/9906070].
}

\lref\tva{T.R. Taylor and C. Vafa, ``RR Flux on Calabi-Yau and
Partial Supersymmetry Breaking'', hep-th/9912152,
Phys.Lett. {\bf B474} (2000) 130-137.}

\lref\mayr{P. Mayr,``On Supersymmetry Breaking in String Theory and its
Realization in Brane Worlds'',
hep-th/0003198, Nucl.Phys. {\bf B593} (2001) 99-126.}

\lref\GiddingsYU{
S.~B.~Giddings, S.~Kachru and J.~Polchinski,
``Hierarchies from fluxes in string compactifications,''
arXiv:hep-th/0105097.
}

\lref\DasguptaSS{
K.~Dasgupta, G.~Rajesh and S.~Sethi,
``M theory, orientifolds and G-flux,''
JHEP {\bf 9908}, 023 (1999)
[arXiv:hep-th/9908088].
}

\lref\curio{
G.~Curio,
``Chiral matter and transitions in heterotic string models,''
Phys.\ Lett.\ B {\bf 435}, 39 (1998)
[arXiv:hep-th/9803224].
}

\lref\CurioAE{
G.~Curio, A.~Klemm, B.~Kors and D.~Lust,
``Fluxes in heterotic and type II string compactifications,''
Nucl.\ Phys.\ B {\bf 620}, 237 (2002)
[arXiv:hep-th/0106155].
}

\lref\SethiES{
S.~Sethi, C.~Vafa and E.~Witten,
``Constraints on low-dimensional string compactifications,''
Nucl.\ Phys.\ B {\bf 480}, 213 (1996)
[arXiv:hep-th/9606122].
}

\lref\BershadskyZS{
M.~Bershadsky, A.~Johansen, T.~Pantev and V.~Sadov,
``On four-dimensional compactifications of F-theory,''
Nucl.\ Phys.\ B {\bf 505}, 165 (1997)
[arXiv:hep-th/9701165].
}
\lref\BershadskyGX{
M.~Bershadsky, A.~Johansen, T.~Pantev, V.~Sadov and C.~Vafa,
``F-theory, geometric engineering and N = 1 dualities,''
Nucl.\ Phys.\ B {\bf 505}, 153 (1997)
[arXiv:hep-th/9612052].
}

\lref\hubsch{
T. Hubsch, {\it Calabi-Yau Manifolds: A Bestiary
for Physicists}, World Scientific 1991.}

\lref\GreenZR{
P.~S.~Green and T.~Hubsch,
``Space-time variable superstring vacua (Calabi-Yau cosmic yarn),''
Int.\ J.\ Mod.\ Phys.\ A {\bf 9}, 3203 (1994)
[arXiv:hep-th/9306057].
}

\lref\bdl{
M.~Berkooz, M.~R.~Douglas and R.~G.~Leigh,
``Branes intersecting at angles,''
Nucl.\ Phys.\ B {\bf 480}, 265 (1996)
[arXiv:hep-th/9606139].
}

\lref\BeckerGJ{
K.~Becker and M.~Becker,
``M-Theory on Eight-Manifolds,''
Nucl.\ Phys.\ B {\bf 477}, 155 (1996)
[arXiv:hep-th/9605053].
}

\lref\MuellerYR{
M.~Mueller and E.~Witten,
``Twisting Toroidally Compactified Heterotic Strings With Enlarged Symmetry Groups,''
Phys.\ Lett.\ B {\bf 182}, 28 (1986).
}

\lref\NarainQM{
K.~S.~Narain, M.~H.~Sarmadi and C.~Vafa,
``Asymmetric Orbifolds,''
Nucl.\ Phys.\ B {\bf 288}, 551 (1987).
}

\lref\AspinwallKE{
P.~S.~Aspinwall,
``A point's point of view of stringy geometry,''
arXiv:hep-th/0203111.
}

\lref\KatzHT{
S.~Katz, D.~R.~Morrison and M.~Ronen Plesser,
``Enhanced Gauge Symmetry in Type II String Theory,''
Nucl.\ Phys.\ B {\bf 477}, 105 (1996)
[arXiv:hep-th/9601108].
}

\lref\KlemmBJ{
A.~Klemm, W.~Lerche, P.~Mayr, C.~Vafa and N.~P.~Warner,
``Self-Dual Strings and N=2 Supersymmetric Field Theory,''
Nucl.\ Phys.\ B {\bf 477}, 746 (1996)
[arXiv:hep-th/9604034].
}

\lref\WittenEM{
E.~Witten,
``Five-branes and M-theory on an orbifold,''
Nucl.\ Phys.\ B {\bf 463}, 383 (1996)
[arXiv:hep-th/9512219].
}

\lref\GiddingsYU{
S.~B.~Giddings, S.~Kachru and J.~Polchinski,
arXiv:hep-th/0105097.
}

\lref\mikerecent{
Shamit Kachru and Michael Schulz, private communication. }

\lref\dasgupta{K. Becker and K. Dasgupta, to appear}

\Title{\vbox{\baselineskip12pt\hbox{hep-th/0208174}
\hbox{SLAC-PUB-9510}
\hbox{SU-ITP-02/35}}} {\vbox{
\centerline{Geometric Constructions}
\medskip
\centerline{of}
\medskip
\centerline{ Nongeometric String Theories}
}}
\bigskip
\bigskip
\centerline{Simeon Hellerman$^1$, John McGreevy$^1$ and Brook Williams$^2$}
\bigskip
\centerline{{\it $^1$ Department of Physics and SLAC, Stanford University,
Stanford, CA 94305}}
\bigskip
\centerline{{\it $^2$ Department of Physics, University of California,
Santa Barbara, CA 93106}}
\bigskip
\bigskip
\noindent
We advocate a framework for constructing perturbative closed string
compactifications which do not have large-radius limits. The idea is to   
augment the class of
vacua which can be described as fibrations by enlarging the monodromy
group
around the singular fibers to include perturbative stringy duality
symmetries.
As a controlled laboratory for testing this program, we study in detail
six-dimensional (1,0) supersymmetric vacua arising from two-torus
fibrations
over a two-dimensional base. We also construct some examples of two-torus
fibrations over four-dimensional bases, and comment on the extension to   
other
fibrations.

\bigskip
\Date{August, 2002}

\newsec{Weakly coupled supersymmetric string vacua without geometry }

In this paper we will examine a new method for
constructing supersymmetric, nongeometric string theories.
The examples on which we focus most closely
make up a class of solutions to supergravity in
$7+1$ dimensions with $32$ supercharges.  These solutions will involve
nontrivial behavior of the metric and Neveu-Schwarz (NS) $B$-field,
but not of any of
the Ramond-Ramond fields, nor of the eight dimensional dilaton.  (The
ten-dimensional dilaton will vary, but only in such a way that the eight
dimensional effective coupling is held fixed.)  We will argue that these
backgrounds are likely to
represent sensible backgrounds for string propagation on which
the dynamics of string worldsheets are determined by a two-dimensional
conformal field theory of critical central charge, with a controlled genus
expansion whose expansion parameter can be made arbitrarily small.

Almost all known examples of perturbative string backgrounds are descibed
by \it nonlinear sigma models\rm, that is, by field theories containing
scalar fields $X\uu\m$
parametrizing a topologically and geometrically nontrivial
target space. The lagrangian for these theories,
\ee{
L\ll{\rm worldsheet} = {1\over{2\pi \alpha^\prime}}
G\ll{\m\n}(X) \pp\ll + X\uu\m \pp\ll - X\uu\n \ \ \ ,
}
describes the dynamics of a fundamental string travelling in
a curved spacetime with metric $G\ll{\m\n}$.  The conditions for
conformal invariance of the worldsheet theory are then
\ee{
0 = \beta\ll{\m\n}[G\ll{\s\t}(X)] = R\ll{\m\n}(X) +
 ({\hbox{\rm quadratic
in }} R\ll{\a\b\g\d}) + \cdots
}
for weak curvatures $R\ll{\a\b\g\d} R\uu{\a\b\g\d} << 1 $
in string units.
Therefore the condition for conformal invariance is approximately the
same as the Einstein equation for the target space metric.  So a
nonlinear sigma model whose target space is a 
smooth Ricci-flat manifold at large volume will always
be approximately conformal, an approximation which improves
if one scales up the manifold $G\ll{\m\n} \to \Lambda G\ll{\m\n}$.

The existence of a large-volume limit of a family of
solutions or approximate solutions gives rise to the {\it moduli problem}:
there are usually several massless scalars in the lower-dimensional
effective field theory corresponding to Ricci-flat deformations of the target
space.  There is always at least one light scalar, or approximate
\it modulus \rm
- namely, the overall size
of the compact space - if flat ten-dimensional
space is an exact solution to the string equations
of motion at the quantum level.  In critical superstring theory, to which
we will restrict our attention exclusively in this paper, flat ten dimensional space
\it is \rm a solution to the equations of motion at the quantum level.

So
the overall volume modulus is an intrinsic difficulty
for geometric compactifications of superstring theory.
Even if potentials are generated for the volume modulus,
supesymmetry forces
such potentials to vanish in the large volume limit, giving rise to
potentials which attract the theory to its least phenomenologically
interesting point \DineHE.

It is therefore important to find compactifications of the theory which lack
an overall scale modulus.  We would like to do this at the perturbative level,
by constructing backgrounds where the closed string worldsheet is controlled
by a 2D conformal field theory which is not a sigma model
and does not have a large volume limit.
To do this, one could in principle scan the space of
two dimensional superconformal field theories of appropriate central charge,
calculate their spectra, and consider only those not connected by
marginal deformations to flat ten-dimensional space; in practice
this would be prohibitively difficult.

Our strategy is to exploit the existence of stringy discrete gauge symmetries of
partially compactified string theory, symmetries which do not commute
with the operation $G\ll{\m\n} \to \Lambda
G\ll{\m\n}$ of rescaling the metric.  There are, indeed, three ways in which
one could in principle exploit such symmetries in order to eliminate the overall
volume modulus:

\item{I.} One could orbifold by them, which would project out the
volume modulus.

\item{II.}  One could simply consider points in the moduli space of
compactifications under which all moduli, including
the volume modulus, are charged; such a vacuum is guaranteed
to be stationary
(though not necessarily stable) against all quantum corrections.

\item{III.} One could consider solutions to string theory with
boundary conditions which lift
the overall volume modulus.

The first possibility has been studied, in the form of the 'asymmetric
orbifold' \refs{\MuellerYR, \NarainQM}, which refers to an orbifold of a torus
by a 'stringy' symmetry, such as $T$-duality, which has no classical
counterpart and under which the volume transforms nontrivially.  The
second possibility has also been proposed, \DineQR, as
a mechanism for solving the moduli problem.  In this paper we investigate
the third possibility.\foot{Some families of solutions of kind (III) will
also contain solutions of kind (I), in
the same way that ordinary Calabi-Yau manifolds often have orbifold
limits at special points in moduli space.}

\subsec{The setup}

The idea is as follows: spacetime is a product of
$10-n$ Minkowski dimensions with an internal
space $X\ll n$.  The internal space is a fiber product of a $k$-dimensional
fiber space $G$ over a $n-k$-dimensional base $\CB$.  Fiberwise,
$G$ solves the string equations of motion (meaning, in the cases we consider,
that it is a flat $T\uu 2$),
and we denote by $\CM$ the moduli space
of $k$-dimensional solutions whose topology is that of
$G$.  We then allow the
moduli $\CM$ of the fiber to vary over $\CB$ in such a way that the total
space $X\ll n$ also solves the equations of motion.
Solving the equations of motion will usually force the lower-dimensional
effective solution to break down at some singular locus $\CS$.  We emphasize
that the microscopic solution generally does \it not \rm break down
at $\CS$.  

We focus on solutions in which, as one circumnavigates the singular locus
$\CS$ of codimension two in $\CB$, the moduli $\CM$ transform
nontrivially under elements of the discrete symmetry group $\CG$ acting on
$\CM$.

The adiabatic approximation in which
$G$ is fiberwise a solution is known in the geometric
context as the semi-flat approximation (\syz).  We will use these
two terms interchangably.  The semi-flat approximation
typically breaks down in the neighborhood
of $\CS$.  Nonetheless when the approximation breaks down, the resulting
singularities can be understood, in all examples we consider,
by using their local equivalence to other known
solutions of string theory.
An analogy which we have found useful is the following:

\centerline{{\it Dirac monopole {\bf :} 't Hooft-Polyakov monopole {\bf ::}
semi-flat metric {\bf:} smooth string background\foot{In the
case of purely geometric monodromies, the 'smooth string background' is
literally a smooth ten-dimensional metric.  In the case of nongeometric
monodromies, there is no smooth metric but the solution is still demonstrably
smooth as a string background.  This can be shown using
perturbative dualities and the known physics of NS fivebranes.} }}

\noindent
The Dirac monopole is a solution to a low-energy
description (Abelian gauge theory)
of a microscopically
well-defined field theory
(for instance $SU(2)$ gauge theory with a Higgs field in the adjoint).
This solution has singular behavior near the core.
For example, its inertial mass (the integrated
Hamiltonian density of the solution) is divergent due to an infinite contribution
from the neighborhood of the origin.
The 't Hooft-Polyakov monopole,
a classical solution of the microscopic theory
with the same charges as the Dirac monopole,
is smooth at the core, but at distances up
to the inverse Higgs vev behaves
just like the Dirac solution.
Had we not known how the Abelian monopole field
should behave near its core, we could
learn this from the 't Hooft-Polyakov
solution.
This is our perspective on the semi-flat approximation --
we use the correct microscopic physics in small patches
when it becomes necessary, and use the semi-flat approximation
to glue these patches together globally.

In all examples in this paper, the fiber $G$ will be a torus, and
we will focus on the case $G=T^2$ as a
controlled laboratory
for testing our ideas.

The organization of the paper is as follows.
In section two, after a brief review
of string theory on $T^2$,
we will derive the effective equations for the metric
and moduli in $\CB$
which express the higher-dimensional equations of motion,
specializing for concreteness to the case of two-dimensional
base.
We then analyze the allowed boundary conditions near the singular
locus $\CS$, and solve those equations locally
in the case
$\CB = \IC$ or $\IC\IP^1$.  We explain the six-dimensional
effective dynamics of these theories involving the degrees of freedom
localized at the degenerate fibers.  In section
three we will construct a global, compact solution.
We deduce its spectrum, and check
its consistency via anomaly cancellation in section four.
Section five presents some alternative descriptions of
this background.
In section six we consider the case where $\CB $ is
four-dimensional.
In section seven we discuss some the many directions for future study.
In several appendices we collect some useful information
about torus fibrations and type IIA supergravity, and study
a generalization of our ansatz.

\newsec{Stringy cosmic fivebranes}

In this section we will find and solve the conditions
for supersymmetric solutions of type IIA string
theory which locally look like compactification on a flat $T^2$ fiber.
We allow the moduli $\tau$ and $\rho$ to vary
over a two-dimensional base $\CB_2$ and
to reach degenerate values on some locus $\CS$.

In bosonic string theory, a restricted class of
such solutions where only $\tau$ {\it or} $\rho$
varies has been known for some time \GreeneYA.
We will try to make clear the connection to
earlier work as we proceed, since
some of our models
have been made using other techniques.
The idea of extending the monodromy
group beyond the geometric one
has been explored previously
in \eg\ \VafaXN, \umanifolds.  The emphasis
in that work was on
finding a higher-dimensional geometric
description of the construction
which we do not require.  In addition,
we have required that our models
have a perturbative string
theory description.

\subsec{Type IIA string theory on $T^2$}

A few basic facts about type II string theory
on a flat two-torus will be useful to us.
We will write the metric on the $T^2$ as
\eqn\twotorusmetric{
M_{IJ} = {V \over \tau_2} \left(
\matrix{
\mid \tau \mid^2 & \tau_1 \cr
\tau_1   & 1
}
\right)_{IJ}.
}
It is convenient to pair the moduli
of the torus into the complex fields
$\tau = \tau_1 + i \tau_2 $ and $\rho = b + i V/2 $
where
$$b \equiv \int_{T^2} B$$
is the period of the NS B-field over the torus.
In eight-dimensional Einstein frame, the relevant part of
the bosonic effective action for these variables is
\eqn\effective{
S = \int_{M_8} d^8 x \sqrt{g} \left( R + {\del_\mu \tau \del^\mu \bar \tau \over \tau_2^2}
                + {\del_\mu \rho \del^\mu \bar \rho \over \rho_2^2} \right)
}

These kinetic terms derive from
the metric
on the moduli space which is invariant under the
\eqn\perturbativeduality{
\CG \equiv O(2,2;\IZ) \sim SL(2, \IZ)_\tau \times SL(2, \IZ)_\rho
}
perturbative duality group, some properties of which we will need
to use.
Under the decomposition indicated in \perturbativeduality,
the first $SL(2, \IZ)$ factor is the geometric modular
group which identifies modular parameters defining
equivalent tori.  The second is generated by
shifts of the B-field through the torus by its period:
$$ b \mapsto b + 1 ,$$
and by T-duality on both cycles
combined with
a $90^\circ$ rotation of the $T^2$.
We will generally use a prime to denote
quantities associated with this second
factor of the perturbative duality group.

One representation of this group comes from its action on
windings and momenta of fundamental strings
on the two-torus.  Labelling these
charges as $w^I$ and $p_I$ for $I=1,2$ along
the two one-cycles, these transform in the $ {\bf 4} = {\bf (2, 1) \otimes (1, 2)}$
vector representation of this group.  By this we mean that they
transform as
\eqn\vectorrep{
\left(\matrix{ p_I \cr \epsilon_{IJ} w^J } \right)
\mapsto
\left(\matrix{ a' \CT & b'\CT \cr
           c' \CT & d'\CT } \right)
\left(\matrix{ p_I \cr \epsilon_{IJ} w^J } \right)
=
\CT \otimes \CT ^\prime
\left(\matrix{ p_I \cr \epsilon_{IJ} w^J } \right)
}
where $\CT = \left(\matrix{a & b \cr c & d}\right)$, $ad-bc=1$.

Another representation of this group is
on Ramond-Ramond (RR) charges.  We
can organize these into
\eqn\rrrep{
\psi \equiv \left(\matrix{ {\rm D2~along~ } \theta_1 \cr
        {\rm D2~ along~} \theta_2 \cr
        {\rm D2~ wrapped~ on~} T^2 \cr
        {\rm D0 } \cr
}\right)
}
which transforms in the reducible Dirac $({\bf 2}, 1) \oplus (1, {\bf 2}) = {\bf 2}_+ \oplus {\bf 2}_-$
representation:
\eqn\diractransf{
\psi \mapsto \left( \matrix{ \CT & 0 \cr
                0 &\CT^\prime} \right) \psi .
}
The periods
$$\left( \matrix{ \omega_1 \cr \omega_2 \cr \upsilon_1 \cr \upsilon_2 }\right) $$
whose ratios are
$$\tau = {\omega_1 \over \omega_2}  ~~~~~ \rho =  {\upsilon_1 \over \upsilon_2} $$
transform in this represenation.

Some further discrete symmetries will be relevant.
We define $\CI_2 $ to be the transformation which inverts the torus:
$$ \CI_2: \theta^I \mapsto - \theta^I .$$
In our notation above, this is $ (\CT, \CT^\prime) = (-1, 1) $.
$\mfl$ reverses the sign of all RR charges, and so can be written as
$ (\CT, \CT^\prime) = (-1, -1) $.  Note that this acts
trivially on the vector (NS) representation as is to be expected.

\subsec{Killing spinors}

The action \effective\ was written in eight-dimensional Einstein frame, but
it will be convenient to study the supersymmetry variations of the
fermions in terms of string frame variables.  We explain
our index and coordinate and frame conventions
in detail in the appendices.  Actually, the results of this
section apply to any $G$-fibration over $\CB$, and
we do not specify the dimension of the fiber or the
base until the next subsection.

The supersymmetry transformations of the gravitino $\Psi\ll {\m\a}$ and
dilatino $\lambda_\a$ in type IIA supergravity in string frame
(see the appendices for details about frames and conventions) are
\eqn\variationofdilatino{
\d \lambda = (\G\ll{[10]}\G\uu\m\pp\ll\m \Phi - {1\over 6} \G\uu{\m\n\s}
H\ll{\m\n\s})\eta
}
\ee{
\d \Psi\ll \m = (\pp\ll\m + {1\over 4}
\Omega_\mu^{ ~\bb {MN} }
~\G\ll{ \bb{MN}}  ) \eta
}
where we have defined the generalized spin
connection $\Omega$ as in \chs\ to be
$$ \Omega_\mu^{ \bb {MN} } \equiv
{\omega\ll\m}\uu{ \bb {MN} } +
{H\ll\m}\uu { \bb {MN} } \G\ll{[10]}
$$
Here we have set to zero RR fields and fermion
bilinears.
$\eta$ is a Majorana but not Weyl spinor of $SO(9,1)$, and
the NS field strength is
\ee{
H\ll{\m\n\s} \equiv (dB)\ll{\m\n\s} \equiv B\ll{\m\n,\s} + {\rm cyclic}
}
\ee{
{H\ll{\m \bb{MN}}} \equiv
H\ll{\m\n\s} E\uu{\n}\ll{\bb{M  }}
E\uu{\s}\ll{\bb{N }}
}
The ten-dimensional chirality matrix is
\ee{
\G\ll{[10]} \equiv {1\over{10!}} \e\ll{\bb{M}\ll 1
\cdots \bb{M} \ll{10}}
\G\uu{\bb {M}\ll 1}\cdots\G\uu{\bb
{M}\ll {10}} =  \G\ll{[10]}\dag.
}

We make an ansatz for the metric $G\ll{\m\n}$, NS two-form $B\ll{\m\n}$
and dilaton $\Phi$ of the form
\eqn\ansatzmetric{
\left [ \matrix {
G\ll{\tilde{\m} \tilde{\n}} & G\ll{\tilde{\m} i} & G\ll{\tilde{\m} J} \cr
G\ll{i \tilde{\n}} & G\ll{ij} & G\ll{i J} \cr
G\ll{I \tilde{\n}} & G\ll{I j} & G\ll{I J}
} \right ]
}
$$=
\left [ \matrix {
\exp{w(x)} \eta\ll{\m\n} & 0 & 0 \cr
0 & g\ll{ij}(x) + M\ll{KL}(x) A\uu K\ll i (x) A\uu L\ll j(x)
& M\ll{JK}(x) A\uu K\ll i (x) \cr
0 & M\ll{IK}(x)  A\uu K\ll j (x) & M\ll{IJ}(x)
}\right ]
$$
\eqn\ansatzbfield{
\left [ \matrix {
B\ll{\tilde{\m} \tilde{\n}} & B\ll{\tilde{\m} i} & B\ll{\tilde{\m} J} \cr
B\ll{i \tilde{\n}} & B\ll{ij} & B\ll{i J} \cr
B\ll{I \tilde{\n}} & B\ll{I j} & B\ll{I J}
} \right ]
=
\left [ \matrix {
0 & 0 & 0 \cr
0 & B\ll{ij}(x) & B\ll{ i J} (x) \cr
0 & B\ll {I  j} (x) & B\ll{IJ}(x)
}\right ]
}
\eqn\ansatzphi{\Phi = \Phi(x)}
where all components of all fields depend only on the coordinates
$x\uu i$ on the base.
This is the semi-flat approximation \refs{\GreeneYA, \syz}.
It is the most general \it ansatz \rm which preserves
a $10-n$-dimensional Poincar\'e invariance and a
$U(1)\uu k$ isometry of the fiber.

For the moment we will further
assume that
\eqn\ansatzabzero{
A\ll i\uu I (x) = B\ll{iI}(x) = 0,
}
which we will relax later on when studying the spectrum
of fluctuations around our solutions.  In the case where
the monodromies are purely geometric, this additional
restriction defines an \it elliptic \rm fibration, in which the
fiber is a holomorphic submanifold of the total space.  In what follows
we willuse the term elliptic more generally to describe the subset of compactifications
where all of the the eight-dimensional vectors are turned off, even in the nongeometric
case.

\subsec{$T^2$ over $\CB_2$}

To our conserved spinor $\eta$ we assign
definite four-, six- and ten-dimensional chiralities
\ee{
\G\ll{[4]} \eta = \chi\ll 4 \eta \hsk \G\ll{[6]} \eta = \chi\ll 6 \eta
\hsk \G\ll{[10]} \eta = \chi\ll {10} \eta
}
\ee{
\G\ll{[4]} \equiv
{1\over {4!}} \e\uu{\bb{ABCD}} \G\uu{\bb{A}} \cdots \G\uu{\bb{D}}
= \G\ll{[4]}\dag
}
\ee{
\Gamma_{[6]} \equiv  \Gamma_{[10]} \Gamma_{[4]} =
 \Gamma_{[4]} \Gamma_{[10]}
}

Next, we notice that the warp factor, $w(x)$, on the
six-dimensional space vanishes given
our ansatz.

The third simplifying observation
is that gamma matrices in
four dimensions satisfy the following identities:
\ee{
\G\uu{\bb{ABC}} = \e\uu{\bb{abcd}} \G\ll{[4]} \cdot \G\uu{ \bb d} =
-  \e\uu{\bb{ABCD}}\G\uu {\bb{D}} \cdot \G\ll{[4]}
}
\ee{
\G\uu{\bb{AB}} \G\ll{[4]} = - \hh \e\uu{\bb{ABCD}} \G\uu{\bb{CD}}
}
%
The vanishing of the dilatino variation
\variationofdilatino\ is
equivalent to
\ee{
\chi_6 \pb \Phi = i V\uu{-1} \pb b
}
and the vanishing of $\d \Psi\ll I$ (where $I$ is an index along
the torus directions) is equivalent to
\eqn\gravitinoalongfiber{
{\O\ll I}\uu {Aa} + \chi\ll 4 \e\uu{AB} \e\uu{ab} {\O\ll I}\uu{Bb}  = 0
}
We can write the generalized spin connection $\Omega$
whose (anti-)self-duality this equation expresses as
\eqn\generalizedconnection{
{\O\ll I}\uu{Aa}
\equiv {\omega \ll I}\uu {Aa} +
\chi\ll{10}
  \e\ll{IJ} f\uu {AJ} e\uu{ai} b\ll{,i}.
}
In appendix D we show that this
implies
\eqn\holcderivreln{
\bar \del \Phi = \bar \del \ln \sqrt V ,
}
which can be solved by setting
\eqn\vandphi{
{V \over e^{2 \Phi } }\equiv g_8^{-2}
}
equal to a constant.
We recognize this undetermined
quantity $g_8$ as the eight-dimensional string coupling\foot{
Note that we would find other solutions for the dilaton
if we turned on the RR potentials.  In that case
we would find F-theory-like solutions
which could remove the zeromode of the dilaton
as well as those of $\rho$ and $\tau$.}.
Given this relationship \vandphi\ between $V$ and $\Phi$,
the remaining equations reduce to
holomorphy of $\rho$ and $\tau$.

We show in appedix D.2 that the
variation of the gravitino with index along the
base will vanish if
the conformal factor on the
base satisfies
\eqn\differenceisharmonic{
0 = \del \bar \del \left( \varphi - \ln \sqrt{ \tau_2 }- \ln \sqrt {\rho_2} \right).
}

\subsubsec{Summary of Killing spinor conditions}

We have found that if $\tau$ and $\rho$ are holomorphic
functions on the base, and the $\rho$ degenerations
back-react on the base metric in the same way
as do the corresponding $\tau$ singularities \GreeneYA,
we preserve six-dimensional $(1,0)$ supersymmetry.
With constant $\rho$ we preserve $(1,1)$, with constant
$\tau$ we preserve $(2,0)$.  The complex structure on
the base is correlated with the six-dimensional
chirality.

\subsec{Stringy cosmic fivebranes}

The insight of \GreeneYA\ is that
our moduli need only be single valued
on the base up to large gauge identifications
in the perturbative duality group $\CG$.
As such, there can be a collection
$\CS$ of branch points in $\CB$ around
which $\tau$ and $\rho$ jump by some
action of $g \in \CG$.
At such a degeneration point, the moduli
must reach values fixed by the element $g$.
Points in the moduli space fixed
by elements of $\CG$ will in general
represent singular tori
or decompactification points.

The basic example of a solution
with nontrivial monodromy is
$$ \tau = {1 \over 2 \pi i} \ln z .$$
In going around the origin, $z \mapsto e^{2 \pi i} z$,
$$ \tau \mapsto \tau + 1$$
which corresponds to the monodromy element
$$ (\CT, \CT^\prime) = \left( \left(\matrix{1&1\cr 0&1} \right), 1 \right) $$
in the notation of \S2.1.

A convenient way to encode the monodromy of $\tau$
is by describing the fiber tori
as a family of elliptic curves
satisfying a Weierstrass equation
\eqn\weierstrass{
y^2 = x^3 + f(z) x + g(z)
}
varying with $z$, a local coordinate on the base.
For this elliptic fibration, the {\it discriminant locus}
where the fiber degenerates
is
\eqn\discriminant{
\CS_\tau = \{ z \in \CB {\rm ~s.t.~} 0 = \Delta_\tau (z) = 4 f(z)^3 + 27 g(z)^2 \} .
}

\subsubsec{Local physics of the degenerations}

In the previous subsections we have shown
that we can preserve supersymmetry
by letting $\rho$ and $\tau$ vary as
locally-holomorphic sections of a $\CG$-bundle.
Branch points around which the
holonomy of the bundle acts
represent degenerations of the
$T^2$ fibration.
The local physics of each possible degeneration
is well-understood, and we review
this understanding in this subsection.

Degenerations of the complex structure $\tau$ of an elliptic
fibration were classified by Kodaira \kodaira\
according to the behavior of the
polynomials in the Weierstrass equation \weierstrass.
This classification (which appears as table I
in the next section) tells us which kind of
singularity
of the total space of the fibration
is created by a particular degeneration of
the fiber.  Assuming supersymmetry
is preserved, these singularities have an
$ADE$ classification.

Note that an $A_0$ singularity is
associated with the degeneration of a particular
one-cycle of the $T^2$ fiber, which
is some integer linear combination $p {\bf a} + q {\bf b}$ of the
a-cycle and the b-cycle.  Therefore,
$A_0$ singularities come with a $\left(\matrix{p\cr q}\right)$
label.

\subsubsec{Local physics of the nongeometric monodromies}

What about degenerations of $\rho$?
Consider a singular fiber near which
\eqn\bfieldshifty{
\rho \sim {N \over 2 \pi i } \ln z .
}
Recalling that $\rho = b + i V/2$, this says that
in going around the origin
the B-field through the torus fiber moves
through N periods,
$$ b \mapsto b + N .$$
We can identify this object
by performing a measurement of the H-flux through
a surface surrounding the singularity.
Such a surface
is the $T^2$-fiber times a circle, $C$, around
the origin, and $H = db \wedge \nu$
where $\nu$ is a unit-normalized volume-form
on the $T^2$, so
\eqn\gaussforhflux{
\int_{T^2 \times C} H =  {N \over 2 \pi i}  \oint_C {dz \over z} = N.
}
This tells us that this homologically-trivial
surface contains $N$ units of
NS5-brane charge \OoguriWJ.  Therefore, we
identify a degeneration of the form \bfieldshifty\
as the semi-flat description of a collection of $N$ type IIA NS5-branes.
Note that this identification is consistent with
the fact that a background with $\tau$ constant
and $\rho$ varying according to \bfieldshifty\
preserves $(2,0)$ supersymmetry,
as does the IIA NS5-brane.

Next we must explain the microscopic
origin of
the other $\rho$ degenerations,
whose monodromies fill out the
rest of $SL(2, \IZ)_\rho$.

For this purpose it is useful to
recall that T-duality along one
cycle of the $T^2$ fiber
replaces type IIA with type IIB.
Choosing different one-cycles
in the $T^2$ along which
to dualize gives different
IIB descriptions of a given
IIA background.  Start with
the NS5-brane associated to the
$\left(\matrix{1 \cr 0 } \right)$ cycle
(whose monodromy is
$$ \left (\CT, \CT^\prime \right) =
\left(1_{2\times 2}, \left(\matrix{ 1&1\cr 0&1} \right) \right) .$$
T-dualizing along the
$\left(\matrix{1 \cr 0 } \right)$ cycle,
one obtains the type IIB $A_0$ \OoguriWJ,
and in particular the one associated with the
$\left(\matrix{1 \cr 0 } \right)$ cycle of the IIB torus.
This is a KK monopole associated with
momentum around this cycle.
If we T-dualize back along the $\left(\matrix{p \cr q } \right)$ cycle
of the type IIB dual torus (\ie\ transverse
to the KK monopole), we obtain
the object in type IIA whose monodromy is
\eqn\volumecutter{
\left(\matrix{ \upsilon_1 \cr \upsilon_2 } \right)
\mapsto
M_{p,q}
\left(\matrix{ \upsilon_1 \cr \upsilon_2 } \right)
}
where $\rho \equiv \upsilon_1 / \upsilon_2$ and
$$  M_{p,q} = \left( \matrix{
    1 - pq & p^2 \cr - q^2 & 1 + pq } \right) $$
is the $SL(2, \IZ)$ matrix which preserves the vector
$\left(\matrix{p \cr q } \right)$.
It is important to notice that this monodromy includes
some action on the volume of the $T^2$ for
any $(p,q)$ other than $(1,0)$.

To what extent is our
semi-flat approximation valid near these objects?
Clearly the actual ten-dimensional solution
for an NS5-brane at a point on a $T^2$ will
break translation invariance on the torus.
In fact, evidence of this
breaking is present
even in the eight-dimensional
smeared solution \GregoryTE.
In particular, Kaluza-Klein momentum
along the torus will
not be conserved because of the H-flux.
The important question
to ask is about couplings.  As we have said, the eight-dimensional
string coupling is arbitrary.  Both the volume
of the fiber, and the ten-dimensional dilaton
diverge near an NS5-brane.
However
tree-level string calculations not involving momentum modes on the torus
are insensitive to this divergence.

\subsec{Putting things on top of other things}

Now that we have discussed the local physics
near each kind of supersymmetric degeneration,
we can consider configurations involving both types,
which we know can preserve only one quarter of
the original supersymmetry.
Special examples of this type, those involving
only $A_N$ degenerations of $\rho$ and arbitrary degenerations
of $\tau$,
are locally described by collections of type IIB
NS5-branes probing ADE orbifold singularities
(\eg\ \BlumFW, \BlumMM, \IntriligatorDH),
a configuration
which is S-dual to D5-branes probing the orbifold
as in \DouglasSW.
However, if the configuration involves
different kinds of degenerations (not just $A_N$) of
both $\rho$ and $\tau$, it cannot be described
in this way.

The metric on the base in
the presence of both $\rho$ and $\tau$ variation
is given by $ds^2 = e^{2 \varphi} \left \abs dz \right \abs^2 $
where the conformal factor is
$$ e^{2 \varphi} = \tau_2  \left \abs
{\eta(\tau(z))^2 \over \Pi_i (z - z_i)^{1/12}
}    \right \abs^2
    \rho_2
\left \abs {\eta(\rho(z))^2 \over \Pi_i (z - \tilde z_i)^{1/12}
}    \right \abs^2
$$
where $\eta$ is the Dedekind eta function.

In the next section we construct and study
in detail a compact model of this kind which
preserves six-dimensional $(1,0)$
supersymmetry.  As we will show,
in a compact model in which both $\tau$ and $\rho$
vary, one must have various
degenerations of each type.

\newsec{Compact models in six dimensions}

What is required to make a compact model
(whose base is a two-sphere with punctures) out of
stringy cosmic fivebranes?
There are two conditions one needs to satisfy.
One is that the conformal factor on the base
behaves smoothly at infinity, i.e.
that the metric on the base is
that of a sphere.
The second condition is that
the monodromy around all of the singularities
be trivial, since this is the same as the monodromy
around a smooth point in $\CB$.

The first condition is easily satisfied
by including the correct number of singularities, weigted appropriately.
(The correct weighting of each singularity for this
purpose counts the order of
vanishing of the discriminant $\Delta$ of the corresponding elliptic curve
~\GreeneYA.)
Since the degenerations of $\rho$
back-react on the metric on the base in the
same way as do the degenerations of $\tau$,
we know from \GreeneYA, \VafaXN\ that this
number is 24.  This is because
equation \differenceisharmonic\
tells us that $\varphi$ is the
two-dimensional electrostatic potential
with charge equal to
the tension $E$ of the degenerations,
which far from the degenerations behaves as
$\varphi \sim - {E \over 2 \pi} \ln \abs z \abs $.
The total tension is $2 \pi \over 12$ times
the total number of times $N$ that the maps
$$ \tau: \IC \to \CF_\tau, ~~~ \rho: \IC \to \CF_\rho $$
cover their fundamental domains, which
says that
far away the metric on the base
looks like
$$ ds^2 = e^{2\varphi} dz d\bar z \sim \abs z^{-N/12} dz \abs^2 ;$$
with $N=24$, the metric behaves as
$$ ds^2 \sim \abs{dz \over z^2}\abs^2 $$
so that infinity is a smooth point
in terms of $u = 1/z$, and we
find $\CB = \IP^1$.

\subsubsec{Trivial monodromy at infinity}

The second condition can be solved in more
than one way.  Including the
action on fermions and RR fields,
it cannot
be solved with fewer
than 12 objects.
For inspiration, we present the following table:

Table 1: Kodaira Classification of Singularities
\bigskip
\begintable
${\rm ord}(f)$ | ${\rm ord}(g)$ | ${\rm ord}(\Delta)$   | monodromy |
                               fiber type | singularity \eltt
$\geq 0$ | $\geq 0$ | $0$ | $\left(\matrix{1&0 \cr 0&1}\right)$ | smooth |none \elt
$0$ | $0$ | $n$ |  $\left(\matrix{1&n \cr 0&1}\right)$ |$I_n$  |$A_{n-1}$ \eltt
$2$ | $\geq 3$ | $n+6$ | $ \left(\matrix{-1&-n \cr 0&-1}\right)$ | $I_n ^*$ | $D_{n+4}$ \elt
$\geq 2$ | $3$ | $n+6$ | $\left(\matrix{-1&-n \cr 0&-1}\right) $|  $I_n ^*$ | $D_{n+4}$ \eltt
$\geq 4$ | $5$ | $10$ |  $\left(\matrix{0&-1 \cr 1&1}\right)$ | $II^*$  | $E_8$\elt
$\geq 1$ | $1$ | $2$| $\left(\matrix{1&1 \cr -1&0}\right)$ |$II$ | none\eltt
$3$ | $\geq 5$ | $9$ |  $\left(\matrix{0&-1 \cr 1&0}\right)$ | $III^*$  | $E_7$\elt
$1$ | $\geq 2$ | $3$ | $ \left(\matrix{0&1 \cr -1&0}\right)$ | $III$ | $A_1$ \eltt
$\geq 3$ | $4$ | $8$ | $\left(\matrix{-1& -1 \cr 1&0}\right)$ |  $IV^*$ | $E_6$ \elt
$\geq 2$ | $2$ | $4$ |  $\left(\matrix{0&1 \cr -1&-1}\right)$ |  $IV$  | $A_2$
\endtable
\bigskip
\noindent
The column labeled ``fiber type'' gives Kodaira's name for the fiber type; the
column labeled ``singularity'' gives the type of singularity
the fiber degeneration causes in the total space.
Note that, as we will clarify below, the contribution that the degeneration makes
to the first chern class of the total space is ${\rm ord} (\Delta) \over 12$.

We have displayed this well-known classification
of degenerations of an elliptic fibration
to point out that the exceptional degenerations come in pairs ($II^*$ and $II$, $III^*$ and $III$,
$IV^*$ and $IV$, $I_0^*$ and $I_0^*$) with monodromies inverse to each other.
These pairs of degenerations all contain 12 simple singular fibers; a space with 12
simple degenerations is asymptotically cylindrical.

This suggests the following construction, which
of course has more concrete descriptions which we give in subsequent
sections.
Choose any pair of $\tau$ degenerations which
has trivial monodromy at infinity, and place them
on a plane; this closes up the plane into a
half-cigar.  Then, independently choose
a pair of $\rho$ degenerations, also with trivial total monodromy,
to make another half-cigar.
The bases of these two two objects may then be glued along 
$S^1$ boundaries, as in the figure below, to make
a compact space.
\ifig\cyl{
We can construct a compact model with both $\rho$ and $\tau$ varying
by gluing together
two asymptotically cylindrical solutions, in each of which
only one varies.
}{\epsfxsize3.0in\epsfbox{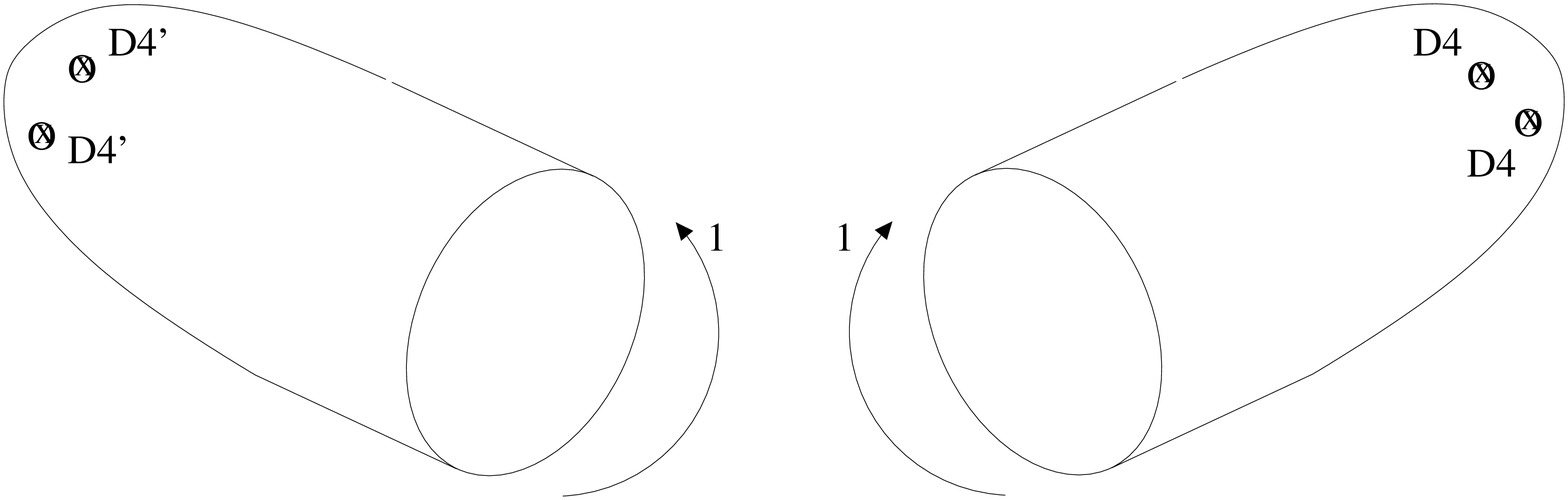}}

It is perhaps most convenient to consider
the example depicted in \cyl\ with two $D_4$
degenerations and two $D_4^\prime$ degenerations\foot{
We denote by $\Gamma^\prime$ the degeneration
of $\rho$ which would be an orbifold singularity $\IC\uu 2/\G$
if it were $\t$ which were degenerating rather than $\r$.
}, since
in that case the values of $\rho$ and $\tau$
can be constant and arbitrary.

\subsubsec{The $12 + 12'$ model}

We can describe the model obtained by gluing these
two cylinders by the following pair
of Weierstrass equations:
\eqn\doubleweierstrass{
y^2 = x^3 + f_{4}(z) x + g_{6}(z) ~~~~
\tilde y^2 = \tilde x ^3 + \tilde f_4(z) \tilde x + \tilde g_{6} (z)
}
by defining
\eqn\taurho{
\tau(z) = j^{-1} \left ( { ( 12 f_4(z) ) ^3 \over 4 f_4(z)^3 + 27 g_6(z)^2 } \right) ~~~~
\rho(z) = j^{-1} \left ( { ( 12 \tilde f_4(z) ) ^3 \over 4 \tilde f_4(z)^3 + 27 \tilde g_6(z)^2 } \right)
}
where $j: {UHP \over SL(2,\IZ)  } \to \IC $
is the elliptic modular function which maps the fundamental
domain for the $SL(2,\IZ)$ action on the upper half plane once
onto the complex plane.
We can think of the first equation of \doubleweierstrass\ as defining the
complex structure of the
actual torus fibers on which we have compactified type IIA; the
second equation determines its complexified K\"ahler form by
specifying the complex structure of the mirror (T-dual along one cycle)
torus.  The degenerations of $\tau$ lie on the locus $\CS_\tau$ of zeros
of $\Delta_\tau$ defined above in eqn. \discriminant\ while the
$\rho$-degenerations lie at the zero locus $\CS_\rho$
of zeros of
$$ \Delta_\rho(z) \equiv  4 \tilde f_4(z)^3 + 27 \tilde g_6(z)^2 .$$

The various points of enhanced symmetry described by
the pairs of degenerations of inverse monodromy in Table 1
can be reached by tuning the polynomials in \doubleweierstrass\
according to the table.  We will, for convenience, count moduli
at a generic point on this coulomb branch, where the zeros of
$\Delta_\tau$ and $\Delta_\rho$ are isolated.

\newsec{The spectrum of the $12 + 12^\prime$ model}

\subsec{``Elliptic'' moduli}

The coefficients of the polynomials $f_4, g_6$ and
$ \tilde f _4 , \tilde g_4$
are moduli of our solution.
$f$ and $g$ contain $5 + 7 = 12 $ coefficients,
of which a rescaling
$$ f_4 \mapsto \lambda^2 f_4, ~~ g_6 \mapsto \lambda^3 g_6 $$
as in \VafaXN\ does not change the torus, leaving $11$
complex parameters.  Similarly
$\tilde f $ and $\tilde g$ give $11$ complex parameters.
Since these are all sections over a single
$\IP^1$, there is one overall $SL(2, \IC)$ action
on coordinates which removes three parameters
leaving $22 -3 = 19$ complex moduli of this kind.

\subsec{RR vectors and tensors in six dimensions}

In this part of the paper we wish to
determine the light fields in
six dimensions that arise
from the ten dimensional RR forms.  We will
do this by reducing these forms along the fiber
as sections of bundles over the base whose
structure group is the monodromy group
$O(2,2)$.  We will then count the
number of such sections in
the semi-flat approximation.  It will be useful
throughout to keep in mind the
example of K3 in the semi-flat approximation.

Reducing the Ramond-Ramond forms on the two-torus fiber,
taking into account their transformation
properties under the perturbative duality group $\CG$,
we learn that
the number of RR tensors and RR vectors in six dimensions
is
\eqn\numbers{
n_T = (b^0_- + b^1_+ + b^2 _-) ~~~ n_V = (b^0_+ + b^1_- + b^2_+).
}
Here
$$b^p_\chi \equiv {\rm dim }_\IR ~ H^p \left(M, V_\chi \right)$$
and $H^p\left(M, V_\chi \right)$ is the $p$th cohomology with
definite eigenvalue $\chi$ of the chirality operator
on the Dirac representation of the $O(2,2;\IZ)$ bundle.
Since from \S2.1 we know that
the $O(2,2)$ chirality, $\chi$, is simply inversion $\CI_2$
of both directions of the torus,
a form has a negative $\chi$ eigenvalue if it has an odd number of
legs on the torus fiber.  Since the Dirac representation
is reducible, we may simply think of $V_\pm$ as
independent rank two bundles.

In the following table, the objects
$1, b, f, \alpha^I, \beta^a$ and wedges thereof
are meant as place-holders
to indicate where the indices of the RR potentials
lie -- respectively, all along the six dimensions, two along the base,
two along the fiber, one along the fiber, one along the
base.
%

Table 2: Reduction of RR fields
\bigskip
\begintable
kind of 6d fields | where to put the indices           | how many fields you get \eltt
scalars | $A_1$ on $\alpha^I$                          | $b^0_-$ \elt
scalars | ($A_1$ on $\beta^a$, $A_3$ on $f \wedge \beta^a$) | $b^1_+$ \elt
scalars | $A_3$ on $b \wedge \alpha^I$                    | $b^2_-$ \eltt
vectors | ($A_1$ on 1, $A_3$ on $f$)                | $b^0_+$ \elt
vectors | $A_3$ on $\alpha^I \wedge \beta^a$            | $b^1_-$ \elt
vectors | ($A_3$ on $b$, $A_7$ on $f \wedge b$)   | $b^2_+$ \eltt
tensors | $A_3$ on $\alpha^I$                      | $b^0_-$ \elt
tensors | ($A_5$ on $f \wedge \beta^a$, $A_3$ on $\beta^a$)   | $b^1_+$ \elt
tensors | $A_5$ on $b \wedge \alpha^I$                     | $b^2_-$
\endtable
\bigskip

We also find the same number $n_T$ of real scalars which
fit into the six-dimensional $(1,0)$ tensor multiplets.
Note that 10-dimensional self-duality of RR fields
relates the 6d scalars in the table to the same number
of four-forms in six dimensions, and relates the
6d vectors to the same number of three-forms in
six dimensions.   However, it relates the tensors to themselves,
rendering them self-dual, as is to be expected of the tensors
in short multiplets of 6d $(1,0)$ supersymmetry.
$$
n_T =  2 \left( h^{(0,0)}_- + h^{(1,0)}_+ \right),
~~ n_V =  2 \left( h^{(0,0)}_+ + h^{(1,0)}_- \right),$$
where $h^{(p,q)}_\pm = {\rm dim}_\IC~(M, V_\pm) $.

\subsec{Index theory on the base}

From the previous subsection we know that
in order to count the spectrum of light RR fields in the
effective six-dimensional theory, we need
to determine dimensions of some
bundle-valued cohomology groups on the base.

The first important point to make is that
because the base is
one complex dimensional,
the number of sections is
almost entirely determined by
the Abelian trace part of the
bundle.
The only effect of the non-Abelian
nature of the structure group
is to reduce the number of
fields by the rank of the representation
except when the bundle is trivial.

We present the answer for the cohomology
which we will justify in what follows.
On the left we have included the results for K3.

Table 3: Cohomology in the semi-flat approximation
\bigskip
\begintable
K3 |            | $12 + 12^\prime$ |  \eltt
$b^0_- = 0 $|$ b^1_- = 20$ |$ b^0_- = 0$|$ b^1 _- = 8 $\elt
$b^0_+ = 2$ | $b^1_+ = 0$ | $b^0_+ = 0$| $b^1 _+ = 8$
\endtable
\bigskip

For K3, one may infer these results from the
fact that there are no RR tensors or scalars
in the spectrum (which implies that
$b^0_- = b^1_+ = 0$), and
from the fact that there are 24 vectors.
In this case the bundle $V_+$ is completely
trivial since the monodromy group
includes no action of $\CT^\prime$.
A trivial rank 2 bundle on $\IP^1$
has two scalar sections and
no one-form sections.  Therefore
$b^0_+ = 2$.
Note that this is consistent with
index theory.
The index theorem for the Dolbeault operator
with values in the representation $R$ of the bundle $V$ says that \EguchiJX
\eqn\dlbltVindextheorem{
\eqalign{
h^{(0,0)}(\CB,V) - h^{(1,0)}(\CB, V) & \equiv {\rm ind}~( \del_V)  \cr =
\int_\CB {\rm Td}(\CB)  {\rm ch}_R (V)
&= \half~{\rm dim} (R) \chi(\CB) + {1 \over 2 \pi} \int_\CB \tr F .
}}
Since the curvature $F_+$ of the bundle $V_+$ is trivial, dim $R = 2$
and
$\chi(\CB) = 2$,
Applied to the Dolbeault
operator with values in $V_+$ this gives:
\eqn\dlbltindexforvplus{
h^{(0,0)}(\CB, V_+) - h^{(1,0)}(\CB, V_+)
=2.
}
Poincar\'e duality
then tells us that $b^2_+ = 2$ as
well and therefore, $b^1_-$ must be
20 to account for the remaining vectors.

We will directly construct the
relevant sections near the $D_4 \oplus D_4$ point, and
then show that our counting is consistent
with $O(2,2)$ index theory.

For definiteness
let us place the singularities at the points $z = 0,1,2,3$.
In the case of K3, all four are $D_4$ singularities;
in our $12+12^\prime$ model, those at $z=0,1$ are $D_4$s,
while those at $z=2,3$ are $D_4^\prime$s.  In
each case, we place the branch cuts run between
$z = 0,1$ and between $z = 2,3$.
The metric on the base is the same in
either case and is
\eqn\metriconbase{
ds^2 = \left \abs dz \over \sqrt{ z (z-1) (z-2) (z-3) } \right \abs^2
}
Note that infinity is a smooth point.
We will count sections of $V_-$ in
each of the two cases; the symmetry
between $\rho$ and $\tau$ in the
$12 + 12^\prime$ model implies
that the number of sections of $V_+$ is the same
in that case.

At this point in the moduli space, it is easy
to ignore the non-Abelian structure, since the
components do not mix under any of the
visible monodromy elements.  Note that this does
not mean that they do not mix locally, close
to the degenerations, and in fact
we will assume that this is the case in
performing our counting.
So we may ignore the fact that the fundamental
representation is actually a doublet,
and, for K3, write a scalar in the
${\bf 2}$ represenation of $V_-$
as
\eqn\scalarminusK{
\psi^{K3} = { h^{K3}_0(z) \over \sqrt{z(z-1)(z-2)(z-3)} } }
where the square roots in the denominator produce
the correct monodromies, and $h^{K3}_0(z)$ is a single-valued function of
$z$.
A holomorphic 1-form valued section of this
bundle is
\eqn\oneformminusK{
\lambda^{K3}
 = { h^{K3}_1(z) dz \over \sqrt{z(z-1)(z-2)(z-3)} } }
where again $h^{K3}_1$ is single-valued.
In the $12 + 12^\prime$ model, the scalar sections of $V_-$ are
\eqn\scalarminus{
\psi = { h_0(z) \over \sqrt{z(z-1)} } }
and the holomorphic 1-form valued sections look like
\eqn\oneformminus{
\lambda
 = { h_1(z) dz \over \sqrt{z(z-1)} } }

Next we must demand that our sections
are nonsingular at the smooth point $z = \infty$.
For the scalar sections, replacing $z$ with
$u = 1/z$, the good coordinate at the other pole,
$$ \psi^{K3}(u) = { h^{K3}_0(1/u) u^2 
\over \sqrt { (1 - u) ( 1 - 2u) ( 1 -3u) }}.$$
So we see that the function $h^{K3}_0(z)$
can grow at most quadratically near infinity.
For the one-form sections, $ dz = - du /u^2$,
so
$$\lambda^{K3} (u) =
{ h^{K3}_1(1/u) du \over \sqrt { (1 - u) ( 1 - 2u) ( 1 -3u) }}$$
which implies that $h^{K3}_1(z)$ can only grow as a constant near infinity.
In the $12+12^\prime$ case we have
$$ \psi(u) = { h_0(1/u) u
\over \sqrt { 1 - u}}$$
so $h_0$ can have at most a simple pole at infinity, while
$$\lambda(u) =
{ h_1(1/u) du \over u \sqrt { 1 - u }}$$
so that $h_1$ must have a {\it zero} at infinity.

The functions $h_{0,1}$ also must behave properly near the
singularities.  At these singularities, the
semi-flat approximation is badly wrong.  However,
as discussed above, the breakdown of the approximation
only occurs in a small region around the
singularity, and we may use our
knowledge of the true {\it local} microscopic physics
to determine conditions on the semi-flat fields
at the boundary of this region.

In this case, our knowledge of the K3 example tells us the
desired information.  Near a $D_4$ singularity,
a scalar-valued section must have at least a simple zero,
and a one-form-valued section can have at most a
simple pole\foot{
Note that the naive norm
on sections (the kinetic term
for the resulting lower-dimensional
fields in the semi-flat approximation)
diverges logarithmically
near the origin for a one-form-valued section
when $h_1$ has a simple pole:
$$ \| \lambda \|^2 = \int_{ \Sigma_0} \lambda \wedge \bar \lambda
= \int_{\Sigma_0} dz d \bar z
{\abs h_1(z)\abs^2  \over \abs z (z-1) \cdots \abs}
\sim \int_0 dr {1 \over r} $$
where $\Sigma_0$ is a small patch around the $D_4$ singularity
at the origin.  This is a breakdown
of the semi-flat approximation exactly analogous
to the divergence of the inertial mass
of the Dirac monopole referred to in the introduction.
We have circumvented this difficulty
by inferring the local physics at the
singularity from well-understood global facts about K3.
}.
For K3, this tells us that $h^{K3}_0(z)$ is
a single-valued function with four zeros that
grows at most quadratically at infinity, and
therefore vanishes.  On the other
hand, the function in the one-form sections on K3 are of the form
\eqn\foneK{
h^{K3}_1(z) = a + {b_1  \over z } + {b_2  \over z -1}
+ {b_3  \over z-2 } + {b_4  \over z -3 }
}
where the five numbers $a, b_1, \dots, b_4$ are complex.
There are therefore ten real $(1,0)$-form-valued sections
of $V_-$, in agreement with Table 3.

This information that we have learned from K3 we
can now apply to the $12+12^\prime$ model.
In this case, $h_0(z)$ must be
a single valued function with two zeros in the complex
plane which grows at most linearly at infinity,
and so must vanish.  The function in the one-form sections
now {\it must} have a pole at one of the $D_4$ singularities
since it must have a zero at infinity and so it is of the form
\eqn\fone{
h_1(z) = { b_1 \over z} +{ b_2 \over z-1 } .
}
Therefore, in the $12 + 12^\prime$ model there are
four independent real sections of this kind,
and we learn that $b^1(\CB, V_-) = 8$.

\subsec{Scalars from NS vectors on the base}

Another interesting component of the spectrum
of this type of model arises from gauge fields
on the base $\CB$ which transform
as sections in the vector
representation of the $O(2,2;\IZ)$ bundle $V$.
One-form-valued zeromodes of such fields
(modulo gauge invariances)
give rise to scalars in six dimensions.
From the killing spinor conditions
including the KK vectors and the
$B_{Ij}$ components, we know that
this zeromode condition is
simply that the connection be flat.

Therefore we can also reduce
this component of the spectrum calculation to
an index problem.
The KK vectors and B-field vectors transform
like string momenta and winding
as in eqn. \vectorrep.  Therefore
${\rm dim}_{\IR} (R) = 4$ and
we should evaluate the first chern class
term in \dlbltVindextheorem\ in the vector
representation.
The number on the left hand side of \dlbltVindextheorem\ is
the number of scalar-valued sections in the vector
representation
minus the number of flat complex Wilson lines
modulo gauge redundancies.

We may evaluate the chern class contribution
to the index by picking a connection on $V$ whose
holonomy induces the desired monodromy
on sections.
Further,
because we know the sections are loclized
near the degenerations, we may
evaluate the contribution of each singularity
by examining only a small patch $\Sigma \subset \CB$ around
the singularity.
Near a $D_4$ singularity at $z=0$, such a connection, in the
vector representation, is
\eqn\dfourconnection{
A = A_z dz = {q \over 2i} {dz \over z} 1_{4 \times 4}
}
where $q$ is an odd integer which is related
to the allowed singular behavior of the sections near
the $D_4$ point.
The connection \dfourconnection\
has holonomy around a curve $C$ surrounding the origin equal to
$$ g_C = e^{ i \oint_C A } = e^{ i \pi q} = - 1_{4 \times 4} $$
as appropriate to a $D_4$ singularity.

The integer $q$ can be determined by appealing to
the results of the previous subsection,
where we have learned that for K3
\eqn\answeranalysis{
- 10 = {\rm ind} (\del_{V_-}) = 2  + 4 c_1(D_4, {\bf 2})
}
where $c_1(D_4, {\bf 2}) = {1 \over 2 \pi} \int_\Sigma {\Tr} F_- $
is the contribution to the first chern class of $V_-$
from a region $\Sigma \subset \CB$ containing
{\it one } of the four $D_4$ singularities.
Hence, $c_1(D_4, {\bf 2})$ turns out to be $-3$.
In this case the relevant connection is
$$ A_- = A_z dz = {q \over 2i} {dz \over z} 1_{2 \times 2} $$
so we learn that
$$ - 3 = {1 \over 2 \pi} \int_\Sigma {\Tr} F_- = q .$$

In the vector representation, the connection appropriate
to a $D_4^\prime$ singularity is the same, since
the monodromies differ by a factor of $\mfl$ which
does not act on the vector representation.  Therefore, the
contribution to the first chern class of {\it either} of these
singularities is equal to
\eqn\chern{
{1 \over 2 \pi} \int_\Sigma \Tr F = {1 \over 2 \pi } \oint_C \Tr A = 2 q
= - 6
}
where $C = \del \Sigma $ is the
boundary of the patch $\Sigma$ containing the singularity
at issue.

Therefore, the total index in the vector
representation for the model with two
$D_4$ singularities and two $D_4^\prime$ singularities
is equal to
\eqn\totalindex{
{\rm ind}~(\del_V) = \half \cdot 4 \cdot 2 - 4 \cdot 6 = - 20.
}
This says that zeromodes of vectors on the base contribute
forty real scalars to the spectrum.
Note that in the analogous calculation for
type II string theory on an elliptic K3, one
also obtains forty real scalars in this manner,
which combine with the $36$ elliptic moduli
plus four more real scalars from the
complexified K\"ahler classes of the fiber and base
to give the $80$ real scalars in the $20$ $(1,1)$
vectormultiplets.

\subsec{Spectrum and anomaly}

Putting together the results of this section,
we find the following spectrum.
In addition to the $8$ tensor
multiplets we found
from RR fields, we find one more
tensor multiplet which contains
the $(1,3)$ piece of the unreduced
NS B-field (the $(3,1)$ piece is
in the six-dimensional $(1,0)$ gravity multiplet),
and whose real scalar is the
six-dimensional dilaton.
We fnd two additional real scalars
from the complexified K\"ahler
class of the base, $\int_\CB (B + i k)$.
Therefore we have found $38 + 40 + 2$
real scalars in addition to the ones in
the tensor multiplets.
This says that we have at a generic
point in the moduli space $n_T = 9$ tensor
multiplets, $n_V = 8$ vector multiplets,
and $n_H = 20$ hypermultiplets.  These
numbers satisfy the condition
for the absence of an
anomaly \GreenBX :
\eqn\anomalycancellation{
n_H - n_V + 29 n_T = 273,}
a highly nontrivial check on the consistency of this background!

\newsec{Other descriptions of these models}

There is a simple sequence of fiberwise
duality operations which relates our models to F theory
models.
Start with our theory in IIA with
$\rho$ and $\tau$ of a $T^2$ varying over
$\CB$. Make the IIA coupling finite
so that, away from branch loci, we may describe it
as M-theory on $S^1 \times T^2 = T^3$ over $\CB$
and perform an 11-9 flip.
The NS B-field $\rho_1$ along $T^2$ maps
to the M-theory three-form along $T^3$.
Further, note that because the ten-dimensional
dilaton is not constant in our solutions,
$R^{11} \propto \sqrt{\rho_2}$.
If we reduce back to IIA along one of the
one-cycles (say the ${\th\uu 1}$-cycle)
of the original $T^2$ we
get a vacuum containing D6 branes at the
degenerations of the torus, the C-field
reduces to an NS B-field with its indices
along 11 and
the ${\th\uu 2}$-cycle of the $T^2$.
T-dualizing along the ${\th\uu 2}$-cycle
gives type IIB with the axion-dilaton
determined by $\tau$; the D6 branes
are replaced by D7 branes.
This theory is compactified
on a $T^2$ whose volume is determined
by the original IIA coupling.
Since the T-duality turns the
NS B-field into shear along this torus,
its complex structure is given by $\rho$.
(The NS B-field through this
torus arises from the wilson line
of the RR 1-form along the ${\th\uu 2}$-cycle
of the original IIA solution, which
lifts in M theory to shear between the
original 11-direction
and the ${\th\uu 2}$-cycle).
Clearly the information derived from our adiabatic
duality can be incomplete
when the semi-flat approximation
breaks down.  This incompleteness will be important when we try
to compare tadpole cancellation in our models and their F-theory duals.

We end up with F-theory
on the doubly-elliptically-fibered
CY over $\CB$ which defined as the
complete intersection \doubleweierstrass\
in a
$\IP^2_{1,2,3} \times \IP^2_{1,2,3}$-bundle
over $\CB$.
In the case of the
$12 + 12^\prime$ model discussed in the previous
section, this duality relates it
to F theory on the doubly-elliptic
Voisin-Borcea
threefold labeled by
$(r, a, \delta) = (10,10,0)$
and $(h^{1,1}, h^{2,1}) = (19,19)$
which was found in \MorrisonPP\ to
have the above spectrum.
Several other six-dimensional models with the
same spectrum have been constructed
\SenTZ, \DabholkarZI.

\subsubsec{Branch structure}

Models with six-dimensional $(1,0)$ supersymmetry
can undergo transitions where
the number of tensormultiplets
decreases by one, and the number
of hypermultiplets increases by 29.
The prototypical example is the
$E_8 \times E_8$ heterotic small-instanton transition
\refs{\oriandami, \sandw}.
Via such a transition,
described in the F theory dual in \MorrisonPP,
it is possible for our model to
reach a large-radius phase.
These tensionless-string transitions
descend to chirality-changing phase
transition in four dimensions
\sande.

\subsec{An asymmetric orbifold description\foot{
We are grateful to Mina Aganagic and Cumrun Vafa for
discussions about such a description.}}

As with many compactifications of string theory,
including Calabi-Yau vacua, the $12 + 12^\prime$ model
has an exactly-solvable point in its moduli space.
In particular, the following asymmetric
orbifold of $T^3 \times R$
realizes a point in the moduli space of the
$12 + 12^\prime$ model:
$$
\alpha: (\theta_1, \theta_2, \theta_3, x) \mapsto
        (-\theta_1, -\theta_2, -\theta_3, -x)
$$
and
$$
\beta: \left\{ \matrix{ (\theta_1, \theta_2, \theta_3, x) \mapsto
       (-\theta_1, -\theta_2, -\theta_3, L - x), \cr
        \Psi_\mu \mapsto \Gamma_{10} \Psi_\mu      ,\cr
    A_{(p)}^{RR} \mapsto - A_{(p)}^{RR} } \right.
$$
More succinctly, $\beta = \mfl
\cdot \CP_L \cdot \CI_4 $, where
$\CP_L$ is a translation by $L$ in the $x$ direction.
The $\theta$s have period $2 \pi$.
The fixed loci of $\alpha$ are (resolved) $D_4$ singularities
at $(\theta_3, x) = (0, 0)$ and at $(\theta_3, x) = ( \pi, 0)$.
The fixed loci of $\beta$ are (resolved) $D_4^\prime$
singularities at
$(\theta_3, x) = (0, L/2)$ and at $(\theta_3, x) = ( \pi, L/2)$.
Note that the group element $\alpha \cdot \beta$ acts
without fixed points.

Using the RNS formalism, one can compute the massless spectrum
of this orbifold.  The calculation is quite similar
to that for the orbifold limit of K3, $T^4 / \vev{\CI_4}$,
the only differences arising from the reversal of
the GSO projection in sectors twisted by $\mfl$.
At each of the eight fixed loci of $\alpha$, the
massless states are the same as in the twisted
sector of $\IR^4 / \vev{\CI_4} $, namely a
vector from the twisted RR sector
and the four real scalars of a
$(1,0)$ hyper from the twisted
NSNS sector.  Each of eight
fixed loci of $\beta$ contribute
the same twisted states as that of
$\IR^4 / \vev{ \CI_4 \mfl} $.  This
consists of
a self-dual tensor and a real scalar
from the twisted
RR sector, and the four real
scalars of a $(1,0)$ hyper from the
twisted NSNS sector.
Unlike $T^4 / \vev{\CI_4}$,
the untwisted RR sector contributes nothing,
while the untwisted NSNS sector is identical
to the symmetric case -- it contains 16
real scalars from
$\psi^i_{-\half} \ket{0}_{NS}^{\rm untw} \otimes
\tilde \psi^j_{-\half} \ket{0}_{NS}^{\rm untw} $
where $i, j$ run over $\theta_i$ and $x$; this
sector also contains the six-dimensional NSNS fields.

Thus we have found the matter content of
the 6d $(1,0)$
gravity multiplet,
8 vectormultiplets, 8 tensormultiplets from RR fields
plus one tensormultiplet containing the dilaton
and half of the NS B-field, and 20 hypermultiplets.
This is anomaly free and agrees with the spectrum computed
in the previous section.

Therefore, one interpretation
of our constructions, at least in this example,
is as a useful description of
deformations of asymmetric orbifolds away
from the orbifold point.

\newsec{$T^2$ over $\CB_4$}


The most immediate way to extend
these ideas beyond
the
six-dimensional $\CN=(1,0)$
models is to enlarge
the base to a space with four real dimensions.
Generically, such solutions will
preserve $\CN=1 $ supersymmetry in four dimensions.
In this section we comment on some
preliminary studies of this
interesting class of backgrounds.
The feature on which we focus our comments is an
eight-dimensional Chern-Simons term which
represents a stringy correction to the action
and BPS equations one obtains from na\"ive reduction of the
ten-dimensional type IIA action.

The BPS equations of the na\"ive 8D supergravity
are satisfied if the base
is a complex twofold with a K\"ahler metric
$g\ll{i\jb}$,
the moduli $\t$ and $\r$ both vary holomorphically over the base, and
\ee{
\det{g\ll{i\jb}} = \sqrt{\rho\ll 2 \tau\ll 2} \cdot ff\st
}
where $f(z\ll 1, z\ll 2)$ is a holomorphic function of the coordinates
on the base.

By the argument given in \S5,
these models are related to
compactifications of F theory on
doubly elliptic fourfolds.  After
an analysis of the local spectrum on the
base and the introduction of some
compact examples, we will
comment further on this duality, and the manner
in which the F theory tadpole
\refs{\SethiES, \BeckerGJ} is cancelled.

\subsubsec{Local spectrum}

Here we discuss the local physics at degenerations
of the semiflat approximation, in type IIA for
concreteness.  With a four-dimensional base,
there are now three interesting classes of degenerate fibers.
First, we can have a degeneration of $\tau$ in complex codimension
one, which is now a complex curve in the base.
%
%
In IIA string theory,
an ADE degeneration of the fiber along a smooth curve of genus
$g$ gives ADE gauge symmetry with $g$ adjoint hypers \KatzHT.
Secondly, $\rho$ can degenerate in complex codimension one;
an ADE NS5-brane wrapping a
smooth genus $g$ curve
leads to $({\rm ADE})\uu g$ gauge symmetry with one adjoint
hyper, as \eg\ in \KlemmBJ.

The new feature in the case when the base is a complex
surface is that these degenerations can intersect.
Intersections of the either the $\tau$ discriminant or
the $\rho$ discriminant with itself
are fairly well understood.
Self-intersections of connected components
of one or the other are included in the definition of 'genus'
used above in stating the generic spectrum.
Intersections of disconnected components give
bifundamental matter \BershadskyZS\ about which more below.

The intersection of the $\rho$ discriminant
and the $\tau$ discriminant is a new
feature of this kind of model.
Matter charged under the gauge
group associated with a $\rho$ singularity along
$\CS_\rho$ descends from objects which
wrap around the one-cycles of $\CS_\rho$ \KlemmBJ.
As a result, at a generic point in the moduli
space (when these one-cycles are not pinched),
we do not find massless matter localized
at these $\rho-\tau$ intersections.
Massive matter charged under both the $\rho$ and $\tau$ groups,
however, is another story, and will in general be present.  Even in
the case where we have a product of $T^2$ with the 4D space we constructed
in section three, there will be regimes of moduli space in which
there are stable brane junctions charged under both groups.

To learn more about these new degenerations
we construct
some instructive compact
examples.

\subsubsec{Compact examples}

A simple class of compact examples can be obtained by taking the base to be
$\CB = \IP^1_1 \times \IP^1_2$.  Choose four asymptotically cylindrical
collections of degenerations and place two
at points on each of $\IP^1_1$ and $\IP^1_2$, wrapping
the other $\IP^1$.
\ifig\alternating{
The most interesting of the examples with $\CB_4 =
\IP^1_1 \times \IP^1_2$ has a $D_4$ and a $D_4^\prime$
wrapping each projective line.
}{\epsfxsize1.5in\epsfbox{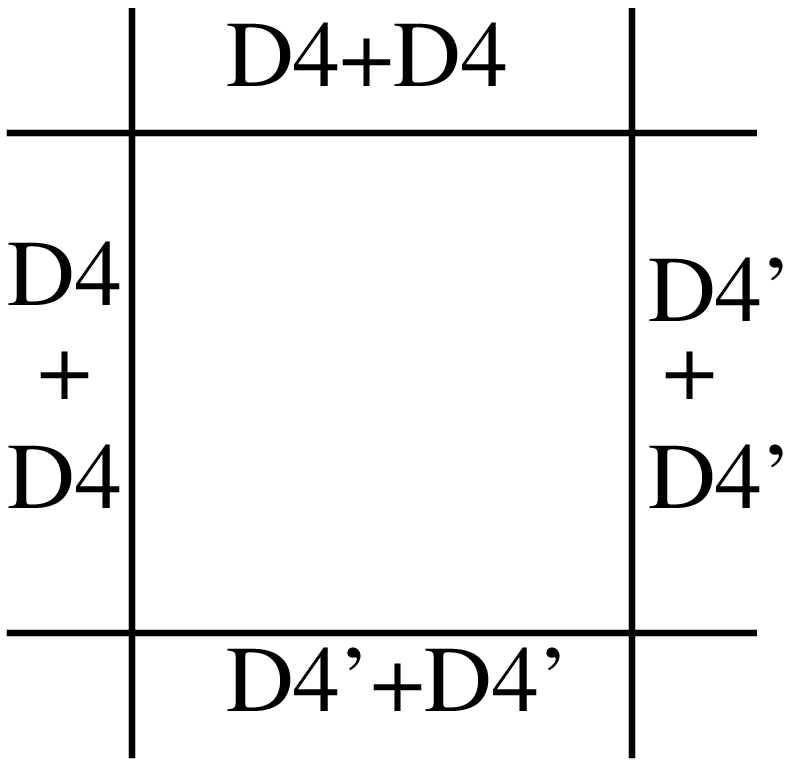}}
If we place all of the $\tau$ degenerations on $\IP^1_1$
and all of the $\rho$ degenerations on $\IP^1_2$, we obtain,
by the duality procedure of \S5,
a model dual to F theory on $K3 \times K3$, which
has enhanced supersymmetry.\foot{An alternative dual description
of this theory, in terms of the heterotic superstring
on a non-K\"ahler
space with $H$-flux, has been brought to our attention \dasgupta.}

These examples fit
into a more general collection
in which
the base is a
Hirzebruch surface $\IF_n$.
We can define the models
as in six dimensions by
a Weierstrass equation
for each of $\tau$ and $\rho$,
\doubleweierstrass\ where
the coefficients now
depend on the coordinates of
the $\IF_n$ base.
One way to be sure that
we satisfy the global
constraints discussed
at the beginning of \S3
for a consistent
compact model is to check
that the dual F theory
model is actually
compactified on a compact Calabi-Yau
fourfold.
Explicitly, this fourfold
is the complete
intersection
\eqn\doubleweierstrassoverbfour{
y^2 = x^3 + f_{4}(stuv) x z^4 + g_{6}(stuv) z^6~~~~
\tilde y^2 = \tilde x ^3 + \tilde f_4(stuv) \tilde x \tilde z^4
+ \tilde g_{6} (stuv) \tilde z^6
}
in the weighted $\IP^2\times \IP^2$ bundle
over $\IF_n$ defined by the
following toric data:
\halign{\indent#\qquad\hfil
&\hfil#\quad\hfil
&\hfil#\quad\hfil
&\hfil#\quad\hfil
&\hfil#\quad\hfil
&\hfil#\quad\hfil
&\hfil#\quad\hfil
&\hfil#\quad\hfil
&\hfil#\quad\hfil
&\hfil#\quad\hfil
&\hfil#\quad\hfil\cr
&$s$&$t$&$u$&$v$&$x$&$y$&$z$&$\tilde x$&$\tilde y$&$\tilde z$\cr
$\lambda$&$1$&$1$&$n$&$0$&$2\alpha$&$3\alpha$&$0$&$2 \tilde \alpha$&$3 \tilde \alpha$&$0$\cr
$\mu$&$0$&$0$&$1$&$1$&$2\beta$&$3\beta$&$0$&$2 \tilde \beta$&$3\tilde \beta$&$0$\cr
$\nu$&$0$&$0$&$0$&$0$&$2$&$3$&$1$&$0$&$0$&$0$\cr
$\rho$&$0$&$0$&$0$&$0$&$0$&$0$&$0$&$2$&$3$&$1$.\cr
}
We need
to specify a chamber for K\"ahler moduli in the quotient
above.  To obtain a positive volume Hirzebruch surface,
we want to have $R_\lambda^2 > n R_\mu^2 > 0 $.
To have geometric torus fibers, we want
$ R_\nu^2, R_\rho^2 > 0$.
Further, the base is only a Hirzebruch surface for $n \leq 12$.

The conditions that \doubleweierstrassoverbfour\ define
a CY fourfold are
\eqn\cycondition{
\alpha + \tilde \alpha = n+2 ~~~~ \beta + \tilde \beta = 2 .
}
We leave a more detailed study of these backgrounds
for future work.

\subsubsec{Tadpoles}

F theory compactifications on
CY fourfolds have tadpoles for the RR
fourform potential polarized along the
four noncompact spacetime directions
\refs{\SethiES, \BeckerGJ}.
The magnitude of this tadpole,
in units of D3-brane charge,
is $1/24$th of the Euler character of the fourfold.
It may be cancelled by placing
space-filling D3 branes at points
on the base of the fourfold,
or by turning on crossed
threeform fluxes on the six-dimensional
base.
The space-filling D3 branes can also be dissolved
into the D7 branes as instantons.


The nongeometric string theories
which we have described do
not have such a tadpole.  This may
be seen, for example, from
the fact that they have an
orbifold limit, analogous to the one given
for the $\CB_2$ case in the previous section.
This is
a consistent (modular invariant)
asymmetric orbifold, discussed in more detail below,
which is connected to the theory by a smooth
deformation of moduli during which
the quantized value of the tadpole
cannot change.
This orbifold limit also makes it clear that
$\CN=1$ supersymmetry is preserved.

As we have emphasized, the intersection
of a $\rho$-singularity with a $\tau$-singularity
is a breakdown of the semi-flat approximation
which is novel in the $\CB_4$ case.
The duality operation which takes us to F theory,
since it uses T-duality on the $T^2$ fiber,
relies on this approximation.
As such, it may be misleading near
points in the base where the approximation
is wrong.
Naive application of this duality tells
us that a $\rho$-$\tau$ intersection maps
to a gravitational instanton on a D7-brane worldvolume
$\CS$,
which would carry (fractional) negative D3-brane charge, due to the coupling
$$- \int_{D7} C_{RR}^{(4)} \wedge {p_1(\CS) \over 24} ,$$
which is one view
of the origin of the
tadpole \SethiES.
We conjecture that the correct map actually
gives such a 7-brane with a real D3-brane
on top of the gravitational instanton, cancelling the
tadpole globally.
Note that such a D3-brane has a branched moduli space,
which includes instanton moduli on the D7-brane,
and coulomb branch moduli which move it in the
two directions transverse to the
D7-brane in the six-dimensional base of the
elliptic fourfold \BershadskyGX.  The orbifold calculation
below provides a check on this idea.

If we move one of these D3-branes to a point $p \in \CB_4$,
away from
special loci ($\CS_\tau$ and $\CS_\rho$)
in $\CB_4$,
we may trust our adiabatic duality map
to determine its image
in the
non-geometric string theory.
Reversing the duality chain of \S5,
we first T-dualize along one of the
one-cycles of the $T^2$ fiber
of the threefold base.  This turns the
D3 brane into a D4 brane wrapping
the ${\th\uu 2}$-cycle of the $T^2$ fiber in
the type IIA background with D6 branes.  Next, we lift to
M theory, and interchange the ${\th\uu 1}$-cycle
with the M-direction.  The D4-brane, which
was not wrapping the ${\th\uu 1}$-cycle,
then becomes an NS5 brane wrapping
the $T^2$ fiber over $p \in \CB_4$.
This then predicts that the
intersection of a $\rho$ degeneration
and a $\tau$ degeneration can
emit a wrapped NS5-brane.
In the generic case the tadpole
is cancelled globally but not locally (see for example
\GiddingsYU\ for a discussion of the F-theory dual to this
effect),
there will be a warp
factor (as in \ansatzmetric) and
varying dilaton on the base,
and the we leave the details
of this for future work.

As further evidence for this picture,
note that the Euler characteristic of a doubly elliptic
fourfold can be rewritten as follows.
Because $\chi(T^2) = 0$,
we have
(see \eg\ \hubsch\ p. 148)
$$\chi(X) = \chi(\CS \subset B_6) $$
where $\CS = \{ p \in B_6 ~{\rm s.t. }~\Delta_\tau(p) = 0 \}$
is the discriminant locus of the $\tau$ fibration.
Because $f,g$ are independent of the coordinates
of the $\rho$ torus ($\tilde x, \tilde y, \tilde z$),
$\CS$ wraps the $\rho$ torus, and therefore is
itself elliptic.
Using the formula for the Euler character of an elliptic
fibration once again,
$$ \chi(X) = \chi(\CS) = \chi( \CS_\rho \cap \CS_\tau \subset \CB_4) $$
where $\CS_\rho = \{ p \in \CB_4 ~{\rm s.t. }~\Delta_\rho(p) = 0 \}$
and $\CS_\tau = \pi(\CS) = \{ p \in \CB_4 ~{\rm s.t. }~\Delta_\tau(p) = 0 \}$
is the image in the four-dimensional base of the $\tau $ discriminant.
But $\CS_\rho \cap \CS_\tau \subset \CB_4 $ is just a collection
of points, so the tadpole (which we interpret as
the number of D3 branes) is
$$ {\chi(X) \over 24} = {1 \over 24} \#
\left(\CS_\rho \cap \CS_\tau \right)_{\CB_4}. $$
This is nothing but
the number of intersection points of the $\rho$-branes
and the $\tau$ branes on the four-dimensional base,
counted with multiplicity and divided by 24.
This number can be measured by the integral formula
\eqn\instantonnumber{
\CI =  C \int_{\CB_4} \omega_\tau \wedge \omega_\rho =
C \int_{\CB_4}  { d \tau \wedge d \bar \tau \over - 8 i \tau_2^2}
    \wedge  { d \rho \wedge d \bar \rho \over - 8 i\rho_2^2} ;
}
this integral counts the number of times the map
\eqn\rhotaumap{
(\tau, \rho): \CB_4 \to \CF \times \CF  }
covers its image\foot{To fix the normalization $C$,
consider the example dual to $K3 \times K3$, where
the map \rhotaumap\ factorizes.  In this case, the Euler character
of the fourfold is $24^2$, which is also equal to
the number of times $\CS_\rho$ intersects $\CS_\tau$.
Therefore
$$ (24)^2 =
\CI(K3 \times K3) = C \int_{\IP^1_1} \omega_\tau \cdot
    \int_{\IP^1_2} \omega_\rho = C \left({2 \pi \over 12} \right)^2 $$
and $C = \left( 24\cdot 12 \over 2 \pi \right)^2 $.  }.

\subsubsec{Asymmetric orbifold limit}

One technique we can use to analyze the spectrum,
and to connect with known models, is to study an orbifold limit.
For concreteness, we
consider the following orbifold, which is in the moduli space of the model
with two $D_4$s and two $D_4^\prime$s on each of two $\IP^1$ factors of the
base, shown in \alternating:
$$\left(T^2 \times S^1 \times \IR \times S^1 \times \IR \right) / \Gamma$$
with coordinates $(\theta^8, \theta^9, \theta^4, x, \theta^6, y)$
(think of $\theta^{8,9}$ as
parametrizing the $T^2$-fiber directions)
where $\Gamma = \vev{ \alpha, \beta, \gamma, \delta} $ and
$$ \eqalign{
\alpha:
&(\theta^8, \theta^9, \theta^4, x, \theta^6, y) \mapsto
(- \theta^8, -\theta^9, -\theta^4, -x, \theta^6, y)   \cr
\beta:
&(\theta^8, \theta^9, \theta^4, x, \theta^6, y) \mapsto
(- \theta^8, -\theta^9, -\theta^4, L -x, \theta^6, y)\otimes(-1)^{F_L}
\cr
\gamma:
&(\theta^8, \theta^9, \theta^4, x, \theta^6, y) \mapsto
(- \theta^8, -\theta^9, \theta^4, x, -\theta^6, -y)   \cr
\delta:
&(\theta^8, \theta^9, \theta^4, x, \theta^6, y) \mapsto
(- \theta^8, -\theta^9, \theta^4, x, -\theta^6, L - y) \otimes(-1)^{F_L}
}
$$
which is very similar to the orbifold in section 5.

Table 4: Massless bosonic spectrum in the asymmetric orbifold limit
\bigskip
\begintable
sector | massless fields     | sector | massless fields \eltt
$\alpha RR$ | 8 complex scalars |$ \alpha NSNS$ | 8 complex scalars \elt
$\beta RR $ | 8 vectors | $\beta NSNS$ | nothing \elt
$\alpha \gamma RR$ | 4 complex scalars |
$\alpha \gamma NSNS$ | 4 complex scalars \elt
$\alpha\delta RR$ | 4 vectors |
$\alpha \delta NSNS$ | nothing \elt
$\gamma RR $ | 8 complex scalars | $\gamma NSNS$ | 8 complex scalars\elt
$\delta RR $ | 8 vectors  |$\delta NSNS$ | nothing \elt
$\beta \delta RR$ | 4 complex scalars |
$\beta \delta NSNS$ | 4 complex scalars\elt
$\beta\gamma RR$ | 4 vectors|
$\beta \gamma NSNS$ | nothing
\endtable
\bigskip
Other sectors $\alpha\beta\gamma$ etc.. contain no massless states.


One point to notice is that
if one modifies $\gamma, \delta$ to something like
$$
\eqalign{
\gamma:
&(\theta^8, \theta^9, \theta^4, x, \theta^6, y) \mapsto
(\epsilon- \theta^8, -\theta^9, \theta^4, x, -\theta^6, -y)   \cr
\delta:
&(\theta^8, \theta^9, \theta^4, x, \theta^6, y) \mapsto
(\epsilon- \theta^8, -\theta^9, \theta^4, x, -\theta^6, L - y)
\otimes(-1)^{F_L}
}
$$
then $\alpha \delta$ no longer has any fixed locus, which means
that at least in the orbifold limit, the spectrum of the $\alpha\delta$
twisted sector cannot be chiral, meaning that if it is ever
chiral, the relevant gauge group is higgsed at the orbifold point.

The fixed locus of $\alpha \delta$ is the intersection of two $D_4 + D_4$s
and two $D_4' + D_4'$s.  Each $D_4$ or $D_4'$ is
made of 12 $\rho$ or 12 $\tau$  branes respectively.
This means that the dual tadpole is cancelled by $4 \times 6 \times 6 / 24 = 12 $
D3 branes.  There are 8 total $\rho-\tau$ intersection
points in the base.  Since 12 is not evenly divisible by 8,
there must be some way for the tadpole-cancelling objects
to fractionate.  This situation is reminiscent of that of \WittenEM.

\subsubsec{Chiral matter}

We can use the
duality with F theory to investigate
chiral matter in these backgrounds.
According to \BershadskyZS, matter in
chiral representations of the geometric
gauge group in F theory on a fourfold
arises along curves $\CC$
of intersection of components of
the discriminant.
Thinking of the components of the
discriminant as collections of
D7-branes in a IIB description,
the
$7-7^\prime$ and $7^\prime-7$
strings are bifundamentals.
Twisting allows the $7-7^\prime$ strings to
have a different number of zero modes
than the $7^\prime - 7$ strings.
This difference is \BershadskyZS\
\eqn\asymmetry{
\chi_{\tau \tau} = 1 - g(\CS) + c_1(\CL)
}
where $\CL$ is a line bundle over
the intersection $\CS$ determined by the
topological twist on the 7-brane worldvolumes.
This spectral asymmetry is just that arising
from branes at angles \bdl\ and as such
$c_1(\CL)$ is related to the angle
of intersection of the two components.

In our models this corresponds to matter
charged under the gauge groups associated to different
components of the $\tau$ discriminant locus.
In the F theory dual, because $f,g$ are independent of
$\tilde x, \tilde y, \tilde z$,
the curve $\CC$ along which they intersect
will always be the
$\rho$-torus.  Thus the chirality
under the $\tau$ gauge group is
simply $\chi_{\tau\tau} = c_1(\CL)$
in these examples.
It seems likely that there is
also interesting matter charged under the
$\rho $ gauge group at a self-intersection
of the $\rho$-discriminant $\CS_\rho $ .

Note that it is not simply any self-intersection of $\CS_\tau$
that leads to chiral matter.  For example, in
the model with $\CB_4 = \IP_1^1 \times \IP_2^1$ and
two geometric $D_4$'s on each $\IP^1$ factor,
there is enhanced supersymmetry; the dual
F theory model has a base which is
the product
$B_6 = T^2 \times \IP_1^1 \times \IP_2^1$.
Clearly the components of the discriminant
along $\IP^1_1$ and those
along $\IP^1_2$ intersect
along $\CC = T^2$, the torus factor of $B_6$.
However,
because the branes intersect
at right angles, there is
no chiral matter in this case.


\newsec{Discussion}

\subsubsec{Fixing the size of the base}

The reader will have noticed that the moduli
of the geometric fibration which are eliminated
by the extension of the monodromy group
are the moduli of the {\it fiber}.  We
have said nothing thus far about fixing the
(complexified) k\"ahler moduli of the base.
However, for this purpose we may
employ another mechanism for fixing moduli,
which is also generically
active in type II string solutions.  This
is the presence of RR fluxes (see \eg\
\refs{\gvw, \tva, \mayr, \GiddingsYU, \DasguptaSS, \CurioAE}).
This mechanism is actually quite
complementary to our
own\foot{We are grateful to Shamit Kachru
for discussions on this point.}.
One finds it difficult \GiddingsYU\
to fix the overall size of the
manifold using fluxes, whereas,
as we have emphasized, nongeometric monodromies
generically eliminate this mode.

\subsubsec{$T^3$ fibrations and junctions of degenerate loci}

From \syz\ we believe that a generic
CY threefold is fibered by special Lagrangian
three-tori.  This fibration degenerates along a real
codimension two locus $\CS$ in the base $\CB$.
Such a locus is a collection of curve segments
joined at junctions \morrison.

In the same way that in the case of $T^2$  fiber
we may think of the back-reaction of the discriminant
locus as closing up the base into a compact manifold,
we should think of $\CS$ of the SYZ fibration as
a {\it gravitating cosmic string junction} -- or more precisely, a
cosmic fourbrane junction, when the noncompact directions are included.

Along each segment, a particular $T^2 \subset T^3$
degenerates, and the monodromy around
this component lies in a $SL(2, \IZ) \subset SL(3,\IZ)$.
In a geometric CY, the total monodromy
around all components of $\CS$ is a subgroup of the
geometric $SL(3, \IZ)$ modular group of the
$T^3$.
More generic solutions will extend
this monodromy group to a larger subgroup
of the duality group, and will preserve
only four supercharges in four dimensions.
Some early work in this direction includes \LiuMB.

\subsubsec{Worldsheet CFT}

As we have emphasized, since the
eight dimensional string coupling is arbitrary,
these models have weak-coupling limits in
which many quantities are computable
using worldsheet conformal field theory
techniques.  Away from orbifold limits,
we have not addressed the
interesting question of which
CFT one should use.
One point to notice is that
the static gauge CFT on a long fundamental string has only
$(0,4)$ worldsheet supersymmetry\foot{
This was emphasized to us by Mina Aganagic.
}.  The worldsheet superalgebra of the RNS superstring prior to
gauge fixing and GSO projection must be $(1,2)$ or $(1,4)$ in order
to allow for a consistent set of superconformal constraints and
reduced spacetime supersymmetry.  We remind the reader
that the
worldsheet SUSY before gauge fixing and GSO projection
and the static gauge SUSY of the long string need not
be the same.  Although the two superalgebras
are the same in the case of CY3
compactification -- namely, they are both
$(2,2)$ -- there is no natural identification between
these algebras, one being (partly) a gauge artifact and the other being a physical
symmetry which acts on the spectrum.  We should not be surprised
that the two superalgebras differ in our case.


%


\subsubsec{Generalizations}

There are two places during the course of our analysis where
our ansatz was not the most general possibility.

One is in appendix E,
where we noticed that it may be possible to preserve
supersymmetry while turning on KK flux
and H-flux in a specific linear combination.
Such a possibility seems to be
related to the mirror image of NS flux.
As a clear missing piece in
our understanding of possible type II flux
backgrounds, this is a subject
of great current interest \mikerecent.

Secondly, we have turned off the RR potentials
in our ansatz.  We found in this ansatz that the volume
of the base and the ten-dimensional IIA dilaton were
related by \holcderivreln.  Turning
on the RR potentials allows the dilaton to be
a constant plus the imaginary part of a holomorphic function
$\sigma$
whose real part is an RR axion.
Another $SL(2, \IZ)$ which
acts by fractional linear transformations on $\sigma$
is obtained by
conjugating $SL(2, \IZ)_\rho$
by an 11-9 flip.
Solutions where the monodromy group
includes action by this $SL(2, \IZ)$
will fix the dilaton, though not at
weak coupling.
By the duality operation described in \S5,
the type IIA dilaton maps to the volume of the
elliptic fiber of the F-theory base, which
means that $\sigma \mapsto -1 / \sigma$ jumps map to
T-dualities in the geometry of the F theory
base.
Therefore a solution in which $\sigma$ jumps
maps to a non-geometric compactification of F-theory!




\subsubsec{Extensions}

We hope we have convinced the reader
that the semi-flat approximation
is a useful physical tool.
When combined
with knowledge about the microscopic
origins of its {\it localized} breakdowns,
it provides an effective handle
on otherwise uncharted solutions.
There is clearly an enormous class of string vacua which
may be described this way; the elliptic
fibrations on which we have focused are
only a special case.
Given a string compactification (not necessarily a geometric one)
whose lower-dimensional effective theory, duality
group, and extended objects are known, one can describe a
large class of less-supersymmetric solutions
with some of the moduli eliminated
using a
a nontrivial fibration of this theory whose monodromies lie in
the full duality group of string theory on the fiber.

Partially geometric constructions of intrinsically stringy vacua
make it possible to disentangle the properties of
these vacua from the obscure machinery of
worldsheet conformal field theory and combine them with such
modern tools as Ramond-Ramond fluxes, branes, and F-theory.  The
prospect of lifting moduli with such a combination offers an inviting
direction for future study.

\vfill\eject

\appendix{A}{Conventions about coordinates and indices}

We choose the following conventions for our coordinates:
$X\uu\m$ are the coordinates on the entire $9+1$ dimensions,
$\tilde{y}\uu{\tilde{\m}}$ are the coordinates on the
$10-n$ Minkowski directions,
$x\uu i$ are the coordinates on the $n-k$-dimensional base,
and $\th\uu I$ are the coordinates on the $T^k$ fiber.

Let $\bb{M},\bb{N},\bb{P},\cdots $ be the entire set of tangent space
indices;
let $a,b,c,\cdots $ be the tangent space indices corresponding to $x\uu i$;
let $A,B,C, \cdots$ be the tangent space indices corresponding to
$\theta\uu I$; let $\bb{A},\bb{B},\bb{C},\cdots $ be indices that run over
both $a$ and $A$.  We take
the $\th\uu I$ coordinates to have constant periodicity $\th\uu I \sim
\th\uu I + n\uu I, n\uu I\in \IZ$.\foot{Yes, the period is $1$ rather than
$2\pi$ even though these coordinates are 'angles' denoted by
'$\theta\uu I$'.  Sorry!}

\appendix{B}{{Conventions for type IIA supergravity}}

In this paper we use ten-dimensional string frame as the starting
point for our conventions.  Here we relate it to \giani.  Before doing
so we restore units and correct some minor errors in the published version
of \giani.

\subsec{Clarifications and Corrections to \rm \giani}

$\bullet$ The Einstein term on page 327 of \giani\ should be multiplied by
$1/k\sqd$.

$\bullet$ The definition of $\sigma$, on page 328, should read
$\sigma \equiv \exp{k\phi / (2\sqrt{2})} = \exp{\sqrt{2} k \phi / 4}$.

$\bullet$
Also note that the sign convention of \giani ~for defining the Ricci
scalar is opposite that of the usual one, in which
\ee{
[\gg\ll \m, \gg\ll\n ] V\uu\s = \Riet {\m \n} \s \t V\uu\t
}

\subsec{Dictionary}

Here we translate between the quantities of \giani, which we denote
with a $\gp$, and our quantities, which in this section we denote
with an $\tp$.

Define

\ee{
k\uu\gp  \equiv {{\sqrt{2} \apr\sqd}\over 2}
}
\ee{
\phi\uu\gp
\equiv {1\over {\alpha^{\prime 2}}} \Phi\uu\tp
}
\ee{
\G\ll{11}\uu\gp \equiv \G\ll{[10]}\uu\tp
}
\ee{
\s\uu\gp  \equiv \exp{\Phi\uu\tp / 4}
}
\ee{
G\uu\gp\ll{\m\n} \equiv
\exp{  - \Phi\uu\tp / 2 } G\uu\tp\ll{\m\n}
}
\ee{
e\uu{\bb{A}\gp}\ll\m \equiv
 \exp{-\Phi\uu\tp / 4} E\uu{\bb {A}\tp}\ll\m
}
\ee{
A\ll{\m\n}\uu\gp \equiv
 - {{2\over{\alpha^{\prime 2}}}} B\ll{\m\n}\uu\tp
}
\ee{
\psi\uu\tp\ll\m \equiv
\tpk \exp{\alpha^{\prime 2} \phi\uu\gp/8}
\left ( \psi\ll\m \uu\gp - {{\sqrt{2}}\over 4} \G\ll{[10]}
\G\ll\m\uu\gp \l\uu\gp
\right )
}
\ee{
\l\uu\tp \equiv
2\alpha^{\prime 2} \exp{- \alpha^{\prime 2} \phi\uu\gp/8}  \l\uu\gp
}
\ee{
\e\uu\gp = \exp{- \Phi\uu\tp/8} \e\uu\tp
}

Upon truncation
to $N=1$ in ten dimensions with a supersymmetry parameter of negative
chirality, our conventions are identical to those of \chs.

We now remind the reader of the transformations of various
geometric quantities under a Weyl transformation
$G\ll{\m\n}\uu\ol
= \exp{2c} G\ll{\m\n}\uu\nw:$
\ee{
E\uu{\bb{A} \ol}\ll\m = \exp{c} E\uu{\bb{A}\nw}\ll\m
}
\ee{
E\uu{\bb{A} \m \ol} = \exp{-c} E\uu{\bb{A}\m\nw}
}
\ee{
\G\uu{\ol\s}\ll{\m\n} = \G\uu{\nw\s}\ll{\m\n} + \d\uu\s\ll\m c\ll{,\n}
+ \d\uu\s\ll\n c\ll{,\m} - G\uu\nw\ll{\m\n} G\uu{\s\t\nw} c\ll{,\t}
}
\ee{
\omega\ll\m\uu{\bb{A B}\ol} = \omega\ll\m\uu{\bb{A B}\ol} +
c\ll{,\n} \left [ E\ll\m\uu{\bb {A}\nw} E\uu{\n {\bb{B} \nw}}
- ({\bb A}\leftrightarrow {\bb B} ) \right ]
}
\ee{
R\uu\ol = \exp{-2c} \left [ R - (D-1) (D-2) (\gg c)\sqd - 2 (D-1)
(\gg\sqd c) \right ]\uu\nw
}
With these definitions and identities,
the IIA action translates from \giani as
follows:
\ee{ L\uu\gp = }
\ee{
- {{e\uu\gp}\over {2k\uu{2\gp}}}R\uu\gp - {{e\uu\gp}\over 2} (\gg
  \phi)\uu{2\gp}
- {{e\uu\gp}\over{12}} \s\uu{-4\gp} (F\ll{\m\n\s}F\uu{\m\n\s})\uu\gp
}
\ee{
= {1\over{\alpha^{\prime 4}}} \sqrt{-{\rm det ~ } G\uu\tp}
\left ( R\uu\tp + 4 (\nabla\phi)\uu{2\tp} - {1\over 3} (H\ll{\m\n\s}
H\uu{\m\n\s})\uu\tp \right )
}
\ee{
= L\uu\tp
}
where we have dropped fermions, $RR$ fields, and total derivatives.
This bosonic $NS$ action
agrees with the action in \chs, minus the terms from the
gauge sector of the heterotic theory.

The supersymmetry variations of the fermions (again, setting to zero
fermion multilinears and $RR$ fields) are:
\ee{
\d\psi\ll\m\uu\tp = \nabla\ll\m\uu\tp \e\uu\tp + {1\over 4}
 (\d\ll\m\uu{[\l\ll 1 }\G\uu{\l\ll 2 \l\ll 3]\tp} )
\G\ll{[10]} \e\uu\tp H\uu\tp
\ll{\l\ll 1 \l\ll 2 \l\ll 3}
}
\eqn\dilatinovariation{
\d\l\uu\tp = - \G\uu{\r\tp} \G\ll{[10]}  \e\uu\tp (\Phi\uu\tp\ll{,\r})
- {1\over 6} \G\uu{\l\ll 1 \l\ll 2 \l\ll 3 \tp} \e\uu\tp
H\uu\tp\ll{\l\ll 1 \l\ll 2 \l\ll 3 }
}

For a gravitino of negative chirality this agrees, as noted before,
with the supersymmetry transformations of \chs.

\appendix{C}{{Geometry of $T^2$ fibrations}}

Now we compute Christoffel symbols given our \it ansatz \rm
\ansatzmetric, \ansatzbfield, \ansatzabzero :
\eqn\\          {
\G\ll{\m\n ;\s} \equiv {1\over 2}( G\ll{\m\s,\n} + G\ll{\n\s,\m} -
G\ll{\m\n,\s})
}

\eqn\\          {
\G\ll{ij;k} = {\G\ll{ij;k}}\uu{[\CB]}  \hsk \G\uu i\ll{jk} =
{\G\uu i\ll{jk}}\uu{[\CB]}
}
\eqn\\          {
\G\ll{ij;I} = 0 \hsk \G\uu I\ll{jk} = 0
}
\eqn\\          {
\G\ll{iI;j} = 0 \hsk \G\uu i \ll{jI} = 0
}
\eqn\\          {
\G\ll{iI;J} = {1\over 2}  M\ll{IJ,i}  \hsk \G\uu I\ll{iJ} = {1\over 2}
M\uu{IK} M\ll{KJ,i} = {1\over 2} M\uu{IK} \bs\gg\ll i M\ll{KJ}
}
\eqn\\          {
\G\ll{IJ;i} = -{1\over 2} M\ll{IJ,i}  \hsk \G\uu i \ll {IJ} = - {1\over 2}
g\uu{ij}  M\ll{IJ,j} = - {1\over 2} \bs\gg\uu i M\ll{IJ}
}
\eqn\\          {
\G\ll{IJ;K} = 0 \hsk \G\uu I\ll{JK} = 0
}

\eqn\\          {
\Riet {\m\n} \s \t \equiv \G\uu\s\ll{\n\t,\m} + \G\uu\s\ll{\m\a}\G\ll{\n\t}\uu\a - \G\uu\s\ll{\m\t,\n} - \G\uu\s\ll{\n\a}\G\ll{\m\t}\uu\a ,
}
defined in such a way that
\eqn\\          {
[\nabla\ll\m, \nabla\ll\n] V\uu\s = \Riet{\m\n}\s\t V\uu\t,
}
where
\eqn\\          {
\nabla \ll\m V\uu \s \equiv \pp\ll\m V\uu\s + \G\ll{\m\t}\uu\s V\uu\t
}

So
\eqn\\          {
\Riet{ij}k l = \riet{ij} k l
}
\eqn\\          {
\Riet{iI}J j = {1\over 2} {(M\uu{-1} \bs\gg\ll i \bs\gg\ll j M)\uu J}\ll I
- {1\over 4} {(M\uu{-1} M\ll{,i} M\uu{-1} M\ll{,j}) \uu J}\ll I
}
\eqn\\          {
\Riet{iI}j J = - {1\over 2} \bs\gg\ll i \bs\gg\uu j M\ll{IJ}
+ {1\over 4} ((\bs\gg\uu j M)M\uu{-1} (\bs\gg\ll i M))\ll{IJ}
}
\eqn\\          {
\Riet{IJ}i j = {1\over 4} ((\bs\gg\uu i M) (M\uu {-1} \bs\gg\ll j M))\ll{JI}
- ({I\leftrightarrow J})
}
\eqn\\          {
\Riet{ij}I J = {1\over 4} {(M\uu{-1} (\bs\gg\ll j M)
M\uu{-1} (\bs\gg\ll i M))\uu I}\ll J - (i \leftrightarrow j)
}
\eqn\\          {
\Riet{IJ}K L = {1\over 4}
{(\bs\gg\uu i M\ll{IL}) (M\uu{-1} \bs\gg\ll i M) \uu K }\ll J
- (I\leftrightarrow J)
}
\eqn\\          {
\Riet{iJ}k l = \Riet{iI} J K = \Riet{ij}k I = \Riet{ij}I k =
\Riet{IJ}K i = \Riet{IJ}i K = 0
}

\appendix{D}{{Derivation of the BPS Equations for the two-dimensional case}}

As our basis of four-dimensional gamma matrices
we take
\eqn\\          {
\G\uu {a = 1} = \s\uu 1 \ot \s\uu 1
~~~~~
\G\uu {a = 2} = \s\uu 2 \ot\s\uu 1
}
\eqn\\          {
\G\uu {A = 1} = \s\uu 3\ot\s\uu 1
~~~~~
\G\uu {A = 2} = 1 \ot \s\uu 2
}
\eqn\\          {
\G\equiv - \G\ll{[4]} \equiv - {1\over 24} \e\uu{{\bb{A} \bb{B} \bb{C}
\bb{D}}} \G\uu{{\bb{A}}} \G\uu{{\bb{B}}} \G\uu{{\bb{C}}} \G\uu{{\bb{D}}}
= 1\ot \s\uu 3
}
Projecting onto spinors $\psi$ of positive chirality $\Gamma \psi =
\psi$, we define
\eqn\\          {
\G\uu{\bb AB} \equiv - {i\over 2} [\G\uu {\bb A} , \G\uu {\bb B}]
}
which, in the basis we have chosen, gives
\eqn\\          {
\G\uu {ab} = \e\uu{ab} \s\uu 3  \hskip .7in \G\uu{AB} = \e\uu{AB}\s\uu 3
}
\eqn\\          {
\G\uu{a = \pm , A = \pm} = \pm 2i \s\uu{\pm}
}
\eqn\\          {
\G\uu{a = \pm , A = \mp} = 0
}
where
\eqn\\          {
V\uu\pm \equiv V\uu 1 \pm i V\uu 2
}
and the Pauli matrices $\s\uu p$ are the usual ones satisfying
$\s\uu p \s\uu q = \d\uu{pq} + i \e\uu{pqr} \s\uu r$.

\subsec{Killing spinor equations along the fiber}

A subalgebra of $SO(4) = SU(2)\times SU(2)$
can only preserve
one spinor if it preserves two, and both must be of the same chirality.
So the condition we want is actually that the covariant derivative
actually annihilate \it two \rm linearly independent
spinors of the same
chirality, $\chi_4$.
Given our ansatz, the condition $\nabla\ll I \psi = 0$ does not involve
the partial derivative $\pp\ll I$ so it only involves
$\OO I {\bb {A}} {\bb {B}}$ which means it only involves
$\OO I a A$ because the other components vanish.
Using the identity
$$ \Gamma^{\bb AB} = \epsilon^{\bb ABCD} \Gamma_{[4]} \Gamma^{\bb CD},$$
the condition is
that the matrix $\OO I a A \G\uu{a A}$ be (anti-) self-dual,
\eqn\alongfiber{
0 = \OO I a A + \chi_4 \epsilon^{AB} \epsilon^{ab} \OO I b B \equiv Q_I^{aA}.
}
Contracting this equation into $f^{AI}$, we learn that
$$ 0 = {2 \over V} \left(
\half e^{ia} \del_i V + \chi_6 \epsilon^{ab} e^{ib} b_i \right) $$
which is equivalent to the Cauchy-Riemann equations
$$ 0 = \bar \del \rho .$$

Contracting \alongfiber\ into $\epsilon^{AC} f^{IC} $ leads
to the same equation.  The only remaining independent part of
\alongfiber\ can be extracted by contracting it with
$f^{BI} V^A V^B $, where $V^A = (1, i) $.  This gives
\eqn\remaining{
0 = Q_I^{aA} f^{BI} V^A V^B =
\half H^I H^J M_{IJ, i} \left( e^{ia} - i \chi_4 \epsilon^{ab} e^{ib} \right)
}
where $H^I \equiv f^{A=1~I} + i f^{A=2~I} $.
Eqn. \remaining\ is equivalent to
\eqn\\          {
0 =  \e\ll{IJ} H\ll J
\left(\del_1 - i \chi_4 \del_2\right)
H\ll I
}
If we choose $\chi_4 = 1$, this is $ 0 = \epsilon_{IJ} H^I \bar \del H^J$.
Since $H\ll I$ can never be zero, is equivalent to
\eqn\\          {
\pp\ll\zb H\ll I = \lambda H\ll I
}
for some function $\lambda(z,\zb)$.  We are working locally so we can always
set $\lambda$ equal to $\pp\ll\zb L$ for some function $L(z,\zb)$, and
so the equation above is solved by
\eqn\\          {
H\ll I = \exp{L(z,\zb)} G\ll I(z)
}
for some function $L$ and holomorphic functions $G\ll I$.  (Remember that
globally neither $L$ nor $G\ll I$ need be single valued!)

We can in fact determine $L$ by using the fact that $\dett f = V$.
We have
\eqn\\          {
V = \dett f = \hh \e\ll{IJ}\e\uu{AB} f\uu A\ll I f\uu B\ll J =
\hh \e\ll{IJ} (f\uu 1\ll I f\uu 2\ll J - f\uu 2\ll I f\uu 1\ll I)
=  {i\over 2} \exp{2 {\rm Re} (L)} \e\ll{IJ} G\ll I G^*\ll J
}
and so we end up with
\eqn\\          {
H\ll I (z,\zb)
= \sqrt{V} \cdot
\left [ - {i\over 2} \e\ll{JK} G\ll J (z) G^*\ll K (\zb) \right ]\uu{-\hh}
\exp{i K(z,\zb)} G\ll I (z)
}
\eqn\\          {
f\uu {A=1}\ll I = {\rm Re} (H\ll I)
~~~~~
f\uu {A=2}\ll I = {\rm Im} (H\ll I)
}
as the most general solution (locally) to the condition that
$\gg\ll I \psi = 0$ for spinors of positive chirality.

Next, notice that changing the quantity $G\ll I$ by
overall multiplication by a holomorphic function $G\ll I \to f(z) \cdot G\ll
I$ changes $H\ll I$ only by a phase $K \to K + {\rm arg} (f(z))$.  So only
the ratio $\t(z) \equiv G\ll 1 / G\ll 2 = H\ll 1 / H\ll 2$ is physical, and
it satisfies the equation $\pp\ll \zb \t = 0$.  It is not difficult to check
that $\t$ is indeed the complex structure $\t$ defined earlier.

\subsec{The other half of the Killing spinor equations}

Having exhausted the content of the equation
$\gg\ll I \psi = 0$, we
turn to $\gg\ll i \psi = 0$.
First we work out the Christoffel symbols, in conformal gauge:
\eqn\\          {
G\ll{ij} = g\ll{ij} = \exp{2\varphi} \delta_{ij}
~~~~~
e\uu i\ll a = \d\uu{ia} \exp{\varphi}.
}
Then
\eqn\\ {
\G\ll{ij}\uu k = (\d\ll{ik} \varphi\ll{,j} + \d\ll{jk}\varphi\ll{,i} -
\d\ll{ij} \varphi\ll{,k} )
}
\eqn\\ {
\oo i a b = \varphi\ll{,j}(\d\uu{ia} \d\uu{jb} - \d\uu{ib} \d\uu{ja})
= \varphi\ll{,j} \e\ll{ij} \e\ll{ab}
}
\eqn\\ {
\oo {i=1} a b + i \oo {i=2} a b = - i \e\uu{ab} \pp\ll\zb \varphi
}
\eqn\\ {
\oo {i=1} a b - i \oo {i=2} a b = + i \e\uu{ab} \pp\ll z \varphi
}
\ee{
\oo i A B = - {1\over {4V}} \e\uu{AB} \e\ll{IJ}
(H\ll I H\st\ll {J,i} + H\st\ll I H\ll{J,i})
}
Parametrizing $H_I$ as \ee { \left [ \matrix{
H\ll 1 \cr H\ll 2 } \right ] = \exp{i K}\cdot
\left ( {V\over{\tau_2}} \right )\uu{1/2}
\left [ \matrix { \t \cr 1 } \right ] , } we get
\ee{
\oo i A B = - {1\over {4\tau\ll 2 }} \e\uu{AB}
( 4 \t\ll 2 K\ll{,i} - 2 {\tau\ll{1,i}})
}
so, using equation for $\tau$ derived in the previous subsection, we have
\ee{
\pp \tau\ll 1 = - i \chi_4 \pp\tau\ll 2
~~~~~
\pb\tau\ll 1 = + i \chi_4 \pb\tau\ll 2
}
\ee{
\oo z A B = \e\uu{AB} \pp\ll z (- K - {i\over 2} \chi_4
{\rm ln} \tau\ll 2 )
}
\ee{
\oo {\zb} A B = \e\uu{AB}
\pp\ll {\zb} (- K + {i\over 2} \chi_4 \ln \tau_2 )}

Defining the generalized spin connection matrix,
$\OOH \mu \equiv \OO \mu {\bb A} {\bb B}
\G\uu{\bb A \bb B}$,
(with $\OO \mu{\bb A} {\bb B} $ as defined in \generalizedconnection )
which enters into the action of the covariant derivative on spinors as
\eqn\\ {
\tilde \gg\ll \m \psi \ll \a \equiv \pp\ll\m \psi\ll\a  -
{i\over 4} (\OOH \m )\ll{\a\b} \psi\ll\b,
}
we have
\eqn\generalizedconnectionmatrix{
\OOH {\zb} \equiv \OOH {i=1} + i \OOH {i=2}
= 2 \left(-i\pp\ll\zb \varphi - {1\over {2V}} \e\ll{IJ} H^*\ll I \pp\ll \zb H\ll J
+ {\bar \del b \over V} \right) \cdot \s\uu 3 ;
}
$\OOH {z}$ is the complex conjugate of this expression.

The condition for the existence of a covariantly constant spinor is then
\eqn\commutatorofderivs{
0 = [\pp\ll z - {i\over 4} \OOH z , \pp\ll\zb - {i\over 4} \OOH \zb ].
}
Using the equation found by imposing the vanishing of the
dilatino variation,
$$ {\bar \del b \over V} = i \chi_6 \bar \del \Phi ,$$
this is
\eqn\\ {
0 = - 4i \pp\ll z \pp\ll\zb \varphi
- {1\over V} \e\ll{IJ} \pp\ll z H^*\ll I \pp\ll\zb H\ll J -
{1\over V} \e\ll{IJ} H^*\ll I \pp\ll z \pp\ll\zb H\ll J
- 4 i \chi_6 \del \bar \del \Phi
}
\eqn\\ {
= - 4i \pp\ll z \pp\ll \zb ( \varphi + \chi_6 \Phi) - {1\over V}  \pp\ll z ( \e\ll{IJ}
 H^*\ll I \pp\ll \zb H\ll J ).
}
Using
\eqn\\ {
\e\ll{IJ} H^*\ll I \pp\ll \zb H\ll J = - 2 i V \pp\ll\zb R
}
where
$$
R \equiv i K - \hh {\rm ln} [-{i\over 2} \e\ll{IJ} G\ll I G^*\ll J].
$$
Rewriting $ \bar \del \Phi = - \half \bar \del \ln \rho_2$,
and noting that $\del \bar \del R = \half \del \bar \del \ln \tau_2$,
the condition \commutatorofderivs\ becomes
\eqn\\ {
0 = \pp\ll z \pp\ll \zb ( \varphi - \half \ln \tau_2 - \half \ln \rho_2 ) .
}

\appendix{E}{Vector modes in the base}

\def\n{\eta_2}

In this appendix, we consider
solutions in which the off-block-diagonal components of the metric
and B-field
are nonzero -- i.e. in which there are 8D vectors turned
on.  This more general ansatz is written in
equations \ansatzmetric\ and \ansatzbfield.
%
The Christoffel symbols and (generalized) spin connection become
\ee{\G\ll{ij;k} = \g\ll{ij;k} + \hh M\ll{IJ,j} A\uu I\ll i A\uu J\ll k
+ \hh M\ll{IJ,i} A\uu I\ll j A\uu J\ll k - \hh M\ll{IJ,k} A\uu I\ll i
A\uu J\ll j}
\ee{+ \hh  M\ll{IJ} (A\uu I\ll{k,i} A\uu J\ll j + A\uu I\ll j A\uu J\ll{k,i}
+ A\uu I\ll{k,j} A\uu J\ll i + A\uu I\ll j A\uu J\ll{k,i}
- A\uu I\ll{i,k} A\uu J\ll j - A\uu I\ll i A\uu J\ll{j,k})}
\ee{ \G\ll{ij;K} = -\hh M\ll{KL} (A\uu L\ll{i,j} + A\uu L\ll{j,i})
-\hh (A\uu L\ll i M\ll{KL,j} + A\uu L\ll j M\ll{KL,i})}
\ee{ \G\ll{iJ;k} = - \hh M\ll{JL} F\uu L\ll{ik} - \hh A\uu L\ll k
M\ll{JL,i} + \hh A\ll i \uu L M\ll{JL,k}
}
\ee{
\G\ll{iJ;K} = \hh M\ll{JK,i}
}
\ee{
\G\ll{IJ;k} = - \hh M\ll{IJ,k}
}
\ee{
\G\ll{IJ;K} = 0
}
\ee{
\G\uu k\ll{ij} = \hh M\ll{IJ} g\uu{kl}
(  A\uu J\ll j {F\uu I}\ll{il} +  A\uu J\ll i F\uu I\ll{jl}) - \hh g\uu{kl}
M\ll{IJ,l} A\uu I\ll i A\uu J\ll j
}
\ee{
\G\uu I\ll{jk} = \hh A\uu{Il} M\ll{JK} (A\uu J\ll j F\uu K\ll{kl} +
A\uu J\ll k F\uu K\ll{jl})
- \hh (\nabla\ll i \uu{\rm [B]} A\uu I\ll j + \nabla\ll j\uu{\rm [B]}
 A\ll i\uu I)
}
\ee{
- \hh M\uu {IJ} (M\ll{JK,k} A\uu K\ll j + M\ll{JK,j} A\uu K\ll k)
- \hh A\uu {Ii}  A\uu J\ll j A\uu L\ll k M\ll{JL,i}
}
\ee{
\G\uu i\ll{jK} = - \hh M\ll{IK} g\uu{il} F\uu I\ll{jl}
+ \hh g\uu{il} A\uu I\ll j M\ll{IK,l}
}
\ee{
\G\ll{jK}\uu I = -\hh M\ll{KL} F\uu L\ll{ji} A\uu{Ii}
+ \hh A\uu L\ll j A\uu {Ii} M\ll{KL,i}
+ \hh M\uu {IJ} M\ll{JK,j}
}
\ee{
\G\uu i\ll{JK} = - \hh g\uu{ij} M\ll{JK,j}
}
\ee{
\G\uu I\ll{JK} = - \hh A\uu{Ii} M\ll{JK,i}
}

\ee{
\oo i a b = {\oo i a b }\uu{\rm [B]} - \hh M\ll{IJ} e\uu {aj} e\uu{bk}
 A\uu I\ll i F\uu J\ll{jk}
}
\ee{
\oo i A a = \hh e\uu {aj} f\uu{AI} ( M\ll{IJ} F\uu J\ll{ij}
- A\uu J\ll i M\ll{IJ,j})
}
\ee{
\oo i A B = \hh f\uu {AI} f\uu B\ll {I,i} - \hh f\uu{BI} f\uu A\ll {I,i}
}
\ee{
\oo I a b = + \hh M\ll{IJ} e\uu {ai} e\uu{bj} F\ll{ij}\uu J
}
\ee{
\oo I A a = + \hh f\uu{AJ} e\uu{ai} M\ll{IJ,i}
}
\ee{
\oo I A B = 0
}

\ee{
\OO i a b = {\oo i a b }\uu{\rm [B]} - \hh M\ll{IJ} e\uu {aj} e\uu{bk}
 A\uu I\ll i F\uu J\ll{jk}
}
\ee{
\OO i A a = \hh e\uu {aj} f\uu{AI} ( M\ll{IJ} F\uu J\ll{ij}
- A\uu J\ll i M\ll{IJ,j} - \chi_{10} H_{Iij})
}
\ee{
\OO i A B = \hh f\uu {AI} f\uu B\ll {I,i} - \hh f\uu{BI} f\uu A\ll {I,i} +
\chi_{10} {\del_i b \over V} \epsilon^{AB}
}
\ee{
\OO I a b = e\uu {ai} e\uu{bj} ( \hh M\ll{IJ} F\ll{ij}\uu J + \chi_{10} H_{Iij})
}
\ee{
\OO I A a = f\uu{AJ} e\uu{ai} (\half M\ll{IJ,i} + \chi_{10} \epsilon_{IJ} \del_i b)
}
\ee{
\OO I A B = 0
}

In the general case, where the one-forms in the base are nonzero,
the BPS equations are satisfied if their field strengths vanish, i.e.
\ee{
H\ll{Iij} = F\uu I\ll{ij} = 0
}
This follows from working out the BPS equations explicitly, but for brevity
we omit these calculations and point out that this fact can be derived
on general grounds.  A spinor field transforms as a scalar under general
coordinate transformations and also trivially under gauge transformations
of the $B$-field.  Since the gauge transformations of the one-forms in the
base are inherited from these local symmetries, it follows that both they
and the (vielbein-contracted) covariant derivatives $\gg\ll a\equiv
E\uu\m_{a} \gg\ll \mu, \gg\ll A\equiv E_A\uu{\m} \nabla_\mu$ are gauge-neutral.
It follows that vectors in the base cannot appear in the BPS equations
except through their field strengths.  With the field strengths set to zero,
then, the BPS equations reduce to the ones we derived with the off-diagonal
components of the metric and B-field set to zero.

From this analysis we conclude that
the BPS equations we found above are unmodified by turning
on KK vectors or off-diagonal modes of the NS B-field
if these have vanishing field strength.  As such,
the analysis presented in \S4 counting zeromodes of these fields
is appropriate.

We also find that
$ 0 = \tilde \nabla _I \psi $
is solved by
\eqn\mirrorofnsflux{
0 = \half M_{IJ} F_{ij}^J + H_{Iij}.
}

The BPS equations also imply the Hermitean Yang-Mills equations
${J\uu k}\ll i {J\uu l}\ll j  F\uu I\ll{kl}  - F\uu I\ll{ij}
= g\uu{ik} {J\uu j}\ll k F\uu I \ll{ij} = 0$ for the
complementary linear combination of the fluxes.  For the case of
a four-dimensional base, these solutions dualize to solutions of type
IIB string theory
with dilaton gradients (a.k.a. 'F theory') and RR fluxes,
of the kind discussed in \GiddingsYU.
It seems likely that by combining these two complementary moduli-fixing
mechanisms, F theory on nongeometric compactifications with RR flux could
fix moduli completely, or at least generate solutions isolated from
any type of large-volume point or dual thereof.

\bigskip
\centerline{\bf{Acknowledgements}}
We would like to thank Mina Aganagic,
Per Berglund, Keshav Dasgupta, Brian Forbes,
Shamit Kachru, Joe Polchinski, Michael Schulz,
Cumrun Vafa for useful
discussions.  SH and JM are grateful to the Harvard theory
group for hospitality while this work was in progress.
BW would like to thank
Gary Horowitz.
The work of SH was supported by the DOE under contract
DE-AC03-76SF00515.
The work of JM was supported in part by National Science
Foundation grant PHY-0097915.  The work of BW was supported
by National Science Foundation grant PHY-0070895.

\listrefs
\end